\documentclass[12pt]{article} 
\usepackage{epsfig}
\usepackage{amsfonts}
\usepackage{latexsym}
\usepackage{amsmath,amssymb}
\usepackage{verbatim}
\usepackage{mathtools}
\usepackage{color}
\usepackage{changebar}
\usepackage[dvipdfm,hypertex]{hyperref}
\def\half{{1\over 2}}
\numberwithin{equation}{section}

 \def\p{\partial}

\newcommand{\bea}{\begin{eqnarray}}
\newcommand{\eea}{\end{eqnarray}}
\newcommand{\be}{\begin{equation}}
\newcommand{\ee}{\end{equation}}
\newcommand{\ba}{\begin{align}}
\newcommand{\ea}{\end{align}}

\newcommand{\W}{\mathcal{W}}
\newcommand{\tr}{\mbox{tr}}

\newcommand{\RR}{\mathbb{R}} 
\newcommand{\ZZ}{\mathbb{Z}} 
\newcommand{\NN}{\mathbb{N}} 

\newcommand{\M}{\mathcal{M}}

  \makeatletter
  \let\over=\@@over \let\overwithdelims=\@@overwithdelims
  \let\atop=\@@atop \let\atopwithdelims=\@@atopwithdelims
  \let\above=\@@above \let\abovewithdelims=\@@abovewithdelims

\DeclareMathOperator{\Tr}{Tr}
\DeclareMathOperator{\diag}{diag}

\begin{document}

\begin{titlepage}

\begin{flushright}
AEI-2011-079\\
NSF-KITP-11-221
\end{flushright}

\vspace{8mm}

\begin{center}
{\LARGE The Gravity Dual of the Ising Model} 

\vspace{8mm}

Alejandra Castro$^a$, Matthias R. Gaberdiel$^b$, Thomas Hartman$^c$, \\ 
Alexander Maloney$^a$, Roberto Volpato$^d$

\vspace{1cm}

{\it $^a$
McGill Physics Department, 3600 rue University, Montreal, QC H3A 2T8, Canada}

\vspace{5mm}

{\it $^b$ Institut f\"ur Theoretische Physik, ETH Zurich, CH-8093 Zurich, Switzerland }

\vspace{5mm}

{\it $^c$ Institute for Advanced Study, School of Natural Sciences,
Princeton, NJ 08540, USA}

\vspace{5mm}

{\it $^d$ Max Planck Institut f\"ur Gravitationsphysik, Albert Einstein Institut, 14476 Potsdam, Germany}
\vspace{1cm}

\begin{abstract}
\noindent

We evaluate the partition function of three dimensional theories of gravity in the quantum regime, where the AdS radius is Planck scale and the central charge is of order one. The contribution from the AdS vacuum sector can -- with certain assumptions -- be computed and equals the vacuum character of a minimal model CFT.  The torus partition function is given by a sum over geometries which is finite and computable.  For generic values of Newton's constant $G$ and the AdS radius $\ell$ the result has no Hilbert space interpretation, but in certain cases it agrees with the partition function of a known CFT.  For example, the partition function of pure Einstein gravity with $G=3\ell$ equals that of the Ising model, providing evidence that these theories are dual.  We also present somewhat weaker evidence
that the 3-state and tricritical Potts models are dual to pure higher spin theories of gravity based on $SL(3)$ and $E_6$, respectively.
\end{abstract}

\end{center}

\setcounter{footnote}{0}

\end{titlepage}
\newpage

\renewcommand{\baselinestretch}{1.1}  
\renewcommand{\arraystretch}{1.5}

\tableofcontents

\section{Introduction}

There are many exactly solvable CFTs in two dimensions, some of which describe important statistical systems.  According to the AdS/CFT correspondence such theories are expected to be dual to theories of quantum gravity in three dimensions.  Given the simplicity of the boundary theory, these dualities, if fully understood, would likely shed light on the nature of holography and the emergence of geometry from quantum field theory.   Potential examples have been studied in the semiclassical regime (see e.g.\ \cite{Witten:2007kt, Maloney:2009ck,Gaberdiel:2010pz}) but so far there is no completely satisfactory example of a fully quantum theory of gravity which is dual to an exactly solvable 2d CFT.

In this paper we will argue that a class of exactly solvable CFTs are dual to certain theories of quantum gravity in AdS${}_3$.  These are strongly coupled gravity theories where the AdS radius is Planck scale.  Nevertheless the path integral of quantum gravity can -- with certain assumptions -- be computed exactly and agrees with that of a known minimal model CFT.  The simplest example is the Ising model, which we conjecture to be dual to Einstein gravity with a particular (Planck scale) value of the cosmological constant.  

Our basic strategy will be to compute the exact gravitational path integral of AdS${}_3$ gravity in Euclidean signature with torus boundary conditions.  This is formally a sum over three-dimensional geometries of the form
\be\label{genzgrav}
Z_{\rm grav}(\tau, \bar\tau) = \int_{\partial {\cal M} =T^2} {\cal D}g \ e^{-S_{E}[g]} \ .
\ee
In this sum over geometries we fix the boundary behaviour of the metric at asymptotic infinity to be a torus with a particular conformal structure, which is labelled by a conformal structure parameter $\tau$. 
The physical interpretation of this path integral is as the finite temperature partition function of quantum gravity in Anti-de Sitter space.  This theory of quantum gravity is, according to the AdS/CFT correspondence, expected to be dual to a boundary conformal field theory.  Thus the path integral \eqref{genzgrav} should equal a CFT partition function at finite temperature.  More precisely, it should equal the CFT partition function on a torus:
\be\label{genzcft}
Z_{\rm cft}(\tau, \bar\tau) = \Tr q^{L_0}\bar{q}^{\bar{L}_0} \ , \quad q = e^{2 \pi i \tau} \ .
\ee
The trace here is over the CFT spectrum.  This means that the gravitational path integral (\ref{genzgrav}) will -- in principle -- compute the complete quantum mechanical spectrum of the dual CFT.

The authors of \cite{Dijkgraaf:2000fq} were the first to interpret a gravitational path integral of the form (\ref{genzgrav}) as a CFT partition function; they dubbed the resulting sum over geometries the ``Black Hole Farey Tail.''  In this computation the bulk theory was a complicated string compactification for which the  path integral  (\ref{genzgrav})  is not exactly computable.

In the search for exactly solvable theories of quantum gravity it is natural to consider simpler theories for which the path integral can be computed and matched with a putative dual CFT.  The simplest candidate gravitational theories are ``pure" theories of gravity which do not contain matter fields: the only degrees of freedom come from the metric, or perhaps some simple generalizations thereof to accommodate supersymmetry or higher spin symmetry.  In these cases the equality $Z_{\rm cft} = Z_{\rm grav}$ imposes a highly nontrivial constraint on the bulk theory, one which appears quite difficult to realize.  
This is because in general there is no clear reason why the sum over geometries (\ref{genzgrav}) should necessarily take the form (\ref{genzcft}) of a Hilbert space trace with positive integer coefficients in the $q,\bar{q}$-expansion. 
For example, it was argued in \cite{Maloney:2007ud} that in the semi-classical regime  the path integral of pure general relativity cannot be interpreted as a Hilbert space trace of the form \eqref{genzcft}.   
In the present paper we will argue that in the strongly coupled regime this is no longer the case: the path integral can be computed and in several cases takes the form of a CFT partition function for certain simple, known CFTs. 

We will start by arguing that the path integral for pure quantum gravity, at any value of the coupling, has the form
\be\label{zgravsum}
Z_{\rm grav}(\tau,\bar\tau) = \sum_{\gamma} Z_{\rm vac}\left( \gamma \tau, \gamma \bar \tau\right) \ .
\ee
Here the sum is over possible topologies, which are labeled by elements $\gamma$ of the group $SL(2,\ZZ)$.  The function $Z_{\rm vac}$ represents the contribution from the vacuum topology.   In the semiclassical limit, this formula can be interpreted as a sum over solutions to the Euclidean equations of motion, where $Z_{\rm vac}$ is the  exponential of the classical action along with an -- in principle infinite -- series of perturbative corrections.  In attempting to calculate (\ref{zgravsum}) we encounter two obstacles.  The first is that the perturbative corrections to $Z_{\rm vac}$ are difficult to compute.  The second is that, at least in the semiclassical limit, the sum over geometries is badly divergent and needs to be regulated in some way.  We will argue that both of these obstacles can be overcome in some cases for pure general relativity and its higher spin generalizations. 

In performing these computations it is crucial that we are studying gravity in a strongly coupled regime where quantum effects are of order one.  To understand why this is important, we recall the fundamental observation of Brown \& Henneaux \cite{Brown:1986nw} that the symmetry group relevant for gravity in AdS$_3$ is the two dimensional conformal group with central charge
\be \label{BHcc}
c = \frac{3\ell}{2G} \ ,
\ee
where $\ell$ is the AdS radius and $G$ is Newton's constant.  This implies that the states of the theory are organized into representations of the (centrally extended) local conformal group.  This statement follows from the symmetry structure of theories of gravity with asymptotically AdS boundary conditions and should be true even in the strongly coupled regime  where $c$ is of order one.  In this regime the representations of the conformal group are highly constrained, unlike the semi-classical case where $c$ is large.  This strongly constrains the possible perturbative contributions to $Z_{\rm vac}$ in (\ref{zgravsum}), as first noted in \cite{Castro:2011ui}.   For example, unlike the semi-classical case, the representations for $c<1$ cannot be unitary unless the central charge is one of the special ``minimal model" values.  The function $Z_{\rm vac}$ is then constrained  to be the vacuum character of a minimal model.  We will proceed by assuming that the expression so obtained for $Z_{\rm vac}$, based on considerations of local conformal invariance, is correct.  This is the simplest  -- and seemingly only self-consistent -- definition of the (\textit{a priori} ill behaved) path integral (\ref{genzgrav}) which is consistent with the symplectic structure of the classical phase space we wish to quantize. 

Remarkably, this will imply that the sum over geometries -- i.e.\ the sum over distinct topologies in (\ref{zgravsum}) -- can be performed explicitly.  The important point is that in these cases the apparently infinite sum over topologies reduces to a {\it finite} sum.\footnote{This is in marked contrast with previous discussions of the black hole Farey tail which involved  infinite sums which need to be regulated.  In those cases it was only possible to obtain a reasonable answer for the regulated sum if the partition function was assumed to be a holomorphic function of $\tau$.  
For supersymmetric theories this can be achieved by considering not the partition function but the elliptic genus, which is automatically holomorphic, as in \cite{Dijkgraaf:2000fq, Kraus:2006nb, Manschot:2007ha, deBoer:2006vg}.  Alternatively one can consider a theory such as chiral gravity for which the partition function is holomorphic by design \cite{Strominger:2008dp, Maloney:2009ck}.  In these cases the sum can be regulated in a simple way using the theory of holomorphic modular forms.  (See \cite{Cheng:2011ay} for a recent related discussion.) In this paper we will not assume holomorphic factorization in the sense of \cite{Witten:2007kt}; we are working directly with the partition sum over real geometries in general relativity.}  The resulting partition function is a modular function built out of minimal model characters. For certain theories of gravity this will agree with the partition function of a known unitary conformal field theory.   

The simplest case where this occurs is for pure Einstein general relativity with $c = {1\over 2}$.  In this case the torus partition function agrees with that of the $c={1\over 2}$ critical Ising model, which is equivalent to a single free fermion.  
Thus we conjecture that the Ising model is dual to pure Einstein gravity at the specified coupling.  This conjecture relies on certain assumptions (spelled out below) about the quantization of pure gravity at strong coupling used to derive (\ref{zgravsum}).

The critical Ising model is but the first of an infinite family of exactly solvable CFTs, the Virasoro minimal models with $c<1$. Virasoro minimal models are labeled by two coprime integers $p<p'$, with central charge 
\be\label{virc}
c(p,p') = 1 - \frac{6(p-p')^2}{pp'} \ .
\ee
The theories with $p'=p+1$, $p>2$, are unitary; the simplest case $(p,p') = (3,4)$ is the Ising model.  The sum over geometries described above also gives the correct partition function of the $(4,5)$ theory, which is the tricritical Ising model.  Thus we conjecture that this CFT is dual to pure gravity with $c = \frac{7}{10}$.  However, when $p>4$ pure gravity is apparently inconsistent. In these cases the partition function $Z_{\rm grav}$ at $c = c(p,p+1)$ cannot be written as an expansion in $q,\bar q$ with positive integer coefficients and a unique vacuum state.  So for these values of the coupling constants pure gravity does not have a consistent, quantum mechanical interpretation in terms of a Hilbert space.

It is therefore natural to ask why the quantum gravity path integral $Z_{\rm grav}$ appears to have a consistent quantum mechanical interpretation only in certain special cases.  A partial answer comes from modular invariance.  CFT partition functions must be modular invariant functions of $\tau$, unchanged under $\tau \to \tau + 1$ and $\tau \to -1/\tau$. In the bulk gravitational theory this is related to general coordinate invariance; the sum over modular images in (\ref{zgravsum}) describes a sum over geometries which are related by large diffeomorphisms.  As we will see, this implies that $Z_{\rm grav}$ and $Z_{\rm cft}$ at $c<1$ are both built from characters of degenerate representations of the Virasoro algebra.  For $p=3,4$ only one modular invariant combination of these characters can be constructed, so it is perhaps not surprising that $Z_{\rm grav} = Z_{\rm cft}$.  
For $p>4$, however, multiple modular invariant combinations exist \cite{Cappelli:1987xt}.  In these cases the partition function $Z_{\rm grav}$ will typically be a non-trivial linear combination of
these modular invariants, and as a consequence will  not equal the partition function of a consistent, unitary CFT.

This apparent failure, however, leads to another possible construction of holographic duals for minimal conformal field theories.  Many exactly solvable CFTs  with $p>4$ have the property that their partition functions are the unique modular invariant constructed out of the characters of
an extended symmetry algebra.  These extended symmetry algebras contain the Virasoro algebra as well as additional fields of higher conformal dimension.
This suggests that we should identify rational CFTs whose partition function is the unique modular invariant of a given chiral algebra and seek a dual theory of gravity based on this algebra.  Following this strategy, we give evidence that the 3-state Potts model, a CFT with central charge $c = \frac{4}{5}$ and chiral algebra $\W_3$, could be dual to a higher spin theory of AdS$_3$ gravity which includes a massless spin 3 gauge field \cite{Blencowe:1988gj}.  This higher spin generalization of three dimensional general relativity can be formulated classically as an $SL(3,\RR)\times SL(3,\RR)$ Chern-Simons theory, which was studied recently in \cite{Campoleoni:2010zq}.  Similarly, the tricritical Potts model with $c = \frac{6}{7}$ could be dual to a novel higher spin theory of gravity based on $E_6$.

We emphasize that the evidence for these dualities comes from the matching of genus one partition functions, which includes all of the spectral data of the CFT but no non-trivial information about correlation functions.   In order to check these dualities one would need to match correlation functions or -- equivalently -- higher genus partition functions, from which OPE coefficients can be extracted by pinching (see e.g.\ \cite{Gaberdiel:2010jf} for a recent discussion).  We will see in section 5 some indications that this is less likely to work for the theories with extended chiral symmetry, namely the 3-state and tricritical Pots models. Thus detailed checks of these conjectures are  desirable.

Ideally, we would like to find the holographic dual of a family of exactly solvable unitary CFTs with a large $c$ limit.  This would allow us to consider a semiclassical limit to obtain the dual of weakly coupled Einstein gravity.  Unfortunately we were not able to find such a family.\footnote{However, it is possible to find families of non-unitary CFTs with bulk duals which admit a semiclassical limit.}  We will consider several families of exactly solvable CFTs, and describe many for which the partition function can indeed be written as a sum over 3d geometries. In some cases we will also identify candidate gravity duals.  A summary of successful cases is presented in table \ref{table:results} in the discussion section of this paper.  This is not an exhaustive search, but rather an assortment of examples of CFTs that may be dual to pure gravity or its higher spin generalizations.

In the next section we derive the general formula (\ref{zgravsum}) for the partition function of three dimensional gravity as a sum over geometries.  In section \ref{s:ising}, the gravity path integral is evaluated at $G = 3\ell$ and identified with that of the Ising model.  Other dualities for pure gravity at $c<1$ are considered in section \ref{s:pureother}.  Similar dualities relating higher spin gravity to CFTs with extended chiral algebras, like the 3-state Potts model, are discussed in section \ref{s:higherspin}.  Finally in section \ref{s:discussion} we conclude with a summary of results and discussion of open questions.  In the appendices we review minimal model CFTs and provide further details on how the sum over modular images in (\ref{zgravsum}) is computed.

\section{AdS$_3$ Quantum Gravity at Strong Coupling}\label{s:gravity}

In this section we discuss the quantization of gravity in the strong coupling regime, where the AdS radius is Planck scale.  We will start by reviewing  known features of the phase space of classical AdS gravity before discussing its quantization.  Although we will not provide a completely rigorous quantization of this phase space, we will argue that the possible answers one can obtain are strongly constrained by the symmetry structure of the theory.  As we will see, this is sufficient to match the gravitational partition function with that of known CFTs for some specific values of the central charge. 

We first review some general features of the path integral in three dimensional quantum gravity with negative cosmological constant, following \cite{Maldacena:1998bw,Dijkgraaf:2000fq,Kraus:2006wn,Witten:2007kt,Maloney:2007ud}.
We wish to compute the Euclidean path integral 
\be\label{g:a}
Z_{\rm grav}=\int_{\partial \M}{\cal D}g \,e^{-cS_E[g]} \ .
\ee
Here the integral is over asymptotically AdS manifolds $\M$ with fixed conformal structure at the boundary.  The classical Euclidean action of a solution is proportional to the central charge $c={3 \ell \over 2 G}$, where $\ell$ is the AdS radius and $G$ is Newton's constant.  We have therefore extracted an explicit factor of $c$ from the Euclidean action $S_E$.
We note that the central charge $c={3 \ell\over 2G}$ is proportional to ${1\over \hbar}$, so it plays the role of the (inverse) coupling constant in the bulk gravity  theory.

In computing this path integral we should sum not just over all metrics on a fixed topological manifold, but also over all possible topologies.  The boundary conditions in \eqref{g:a} fix the topology of spacetime only at the boundary, not in the interior.  Thus the gravitational path integral should take the form
\be\label{g:aa}
Z_{\rm grav} = \sum_{\cal M} Z({\cal M}) \ ,
\ee
where $\cal M$ denotes a particular topological three-manifold and $Z({\cal M})$ the contribution from the sum over all metrics on $\cal M$.  

We are now faced with the question of which manifolds $\cal M$ contribute to the sum \eqref{g:aa}.
In the semi-classical approximation, this question can be answered as follows.  The partition function should be expanded in the saddle point approximation as a sum over all solutions to the equations of motion
\be\label{g:b}
Z_{\rm grav}=\sum_{g_{cl}}\exp\left(-cS_E[g_{cl}] + S^{(1)}[g_{cl}] + \frac{1}{c}S^{(2)}[g_{cl}] + \cdots \right)~.
\ee
Here $g_{cl}$ is a classical saddle, i.e.\ a solution of the equations of motion, and $S_E$ is the classical action.  The other terms describe loop contributions, which are suppressed by powers of the coupling constant (the central charge) $c$.  The result is that in the semiclassical limit the only topologies which contribute to the path integral are those which admit a classical solution to the equations of motion.

We now wish to consider the quantization of the strongly-coupled theory, where $c$ is of order one.  Our central assumption is that the conclusion described above -- that the only topologies which contribute are those which admit a classical solution -- continues to hold.  Thus we are requiring that the path integral should be organized as a sum over classical solutions, along with an (in principle infinite) series of quantum corrections around each saddle.  We assume this even though the coupling constant is of order one, so that there is no sense in which the loop corrections are small compared to the classical  Euclidean action. 

This assumption can be viewed as part of our definition of the formal sum over geometries.  Indeed, quantum field theory path integrals can typically only be defined in terms of a sum over classical solutions along with a series of perturbative corrections around each classical solution.  There are even cases where the path integral can be computed exactly using other methods, and shown to agree precisely with a sum over classical saddles, each dressed by a series of quantum corrections.  In these cases the saddle point approximation is, once one includes all perturbative corrections, exact.  A notable example where this is the case is Chern-Simons gauge theory (see e.g.\ \cite{LR}).
Given the similarity between three dimensional gravity and Chern-Simons theory, it is not unreasonable to hope that the same may be true here.
Without a precise definition for the path integral of quantum gravity, however, this should be regarded as a conjecture, albeit a very natural one.

We will now discuss the sum over topologies, before turning to evaluate the contribution from a fixed topology. 

\subsection{Sum over topologies}

Quantum gravity in Anti-de Sitter space makes sense only if one fixes the asymptotic boundary conditions on the metric appropriately.  In AdS${}_3$ the metric is required to approach that of a two-manifold with fixed conformal structure.  We now wish to organize the path integral \eqref{g:b} as a sum over topologies with fixed boundary conditions,  following \cite{Maldacena:1998bw,Dijkgraaf:2000fq, Maloney:2007ud}.

The integral we will perform is over Euclidean 3-manifolds $\cal M$ whose conformal boundary is a two-torus.  
The simplest geometry $\cal \M$ which contributes to this sum is thermal AdS$_3$, which is a Euclidean geometry with metric
\be\label{g:tads}
{ds^2\over \ell^2}=d\rho^2+\cosh^2\rho\, dt_E^2+\sinh^2\rho\, d\phi^2~
\ee
where the angle $\phi$ is periodic $\phi\sim\phi+2\pi$.
This is the geometry obtained by continuing the usual Lorentzian AdS metric in global coordinates to Euclidean signature by taking $t \to t_E = i t$.
Defining the complex coordinate $z=i(\phi+it_E)$, we see that at the boundary $\rho \to \infty$ the metric approaches the usual flat metric $dz d {\bar z}$.  The coordinate $z$ is identified according to 
\be
z~\sim~ z +2\pi i n,~~~~~n \in \ZZ
\ee
so that this is the flat metric on the cylinder.  In order to define thermal AdS we must impose an additional identification of the $z$ coordinate, which we will write as
\be\label{g:baa}
z ~ \sim ~z +  2\pi i m\tau, ~~~~~m\in \ZZ \ ,
\ee
where $\tau$ is a complex parameter.
This is the usual Euclidean identification that defines a grand canonical partition function at finite temperature and angular potential.
  The boundary is now a torus, whose conformal structure is determined by the parameter $\tau$.
 
Topologically, thermal AdS${}_3$ is a solid torus. 
In this solid three-geometry one of the cycles of the boundary torus is contractible and the other is not.  In the case of thermal AdS${}_3$, the contractible cycle is the spatial $\phi$ circle.   
This makes it clear how one should construct other geometries that will contribute to the partition function.
One just needs to consider other solid tori where other cycles of the boundary torus are contractible in the interior.  Geometrically, these new solid tori will have the same metric as thermal AdS, but will be glued on to the boundary torus in a different way. That is to say, they will be related to thermal AdS by a large diffeomorphism which acts non-trivially on the boundary torus.  Such large diffeomorphisms, known as modular transformations, will lead to distinct contributions to the gravitational path integral.  

Let us illustrate this with a simple example. Consider the diffeomorphism
\be
z \to z'=i(\phi'+it_E')=-{1\over \tau}z~.
\ee
The periodicities \eqref{g:baa} transform as
\be
z'~\sim~ z' +2\pi i n- 2\pi i m{1\over\tau }~,\quad n,m\in \ZZ~.
\ee
Thus $\tau$ has been transformed to  $-{1\over \tau}$. After making this change of variables, the contractible cycle is now a combination of $t_E'$ and $\phi'$.  The $\phi'$ cycle is not contractible, and thus the Lorentzian continuation of this geometry will have a horizon. Indeed this geometry is the Euclidean continuation of the BTZ black hole of \cite{Banados:1992wn}. 
 
More generally, the group of non-trivial large diffeomorphisms (i.e.\ the mapping class group) of the boundary torus is the modular group $SL(2,\ZZ)$.  These modular  transformations  act on the conformal structure as
\be\label{g:ba}
\tau \to \gamma\tau ={a \tau + b\over c\tau + d}~, \qquad \gamma = \left({a~~b\atop c~~d}\right) \in SL(2,\ZZ)~.
\ee
We emphasize that although these geometries are related by diffeomorphisms they are physically inequivalent.  Only diffeomorphisms that act trivially at infinity are true gauge symmetries.  Diffeomorphisms that act non-trivially at infinity generate new states in the theory and lead to new contributions to the path integral.  For example, a modular transformation changes the identification of space and time, and will therefore change physical quantities such as the mass and angular momentum of a particular solution. 
 Each element of $SL(2,\ZZ)$ will (up to an equivalence described below) lead to a topologically distinct solid torus, giving rise to a family of Euclidean solutions \cite{Maldacena:1998bw}. 
 
It is possible to show that the geometries described above are the only smooth solutions to the equations of motion with torus boundary conditions (see e.g.\ \cite{Maloney:2007ud} for a simple proof).\footnote{We assume here that only smooth manifolds contribute to the path integral.  We will not consider geometries which admit orbifold-type singularities, nor will we consider complexified geometries where the metric is not real.  In order to address these questions precisely we would need a better understanding of the integration contour for the path integral of general relativity.} Thus the geometries are labelled by the elements $\gamma$ of the group $SL(2,\ZZ)$.  We will refer to the contribution to the partition function of  thermal AdS$_3$ as $Z_{\rm vac}(\tau,\bar\tau)$. All of the other contributions to the path integral are obtained by modular transformations, so that  
\be
Z_{\gamma}(\tau,\bar\tau)=Z_{\rm vac}(\gamma\tau, \gamma\bar\tau)~, \quad \gamma\in SL(2,\ZZ)~.
\ee
Here $Z_{\gamma}(\tau,\bar\tau)$ is the contribution to the path integral from the geometry related to thermal AdS$_3$ by the diffeomorphism which takes $\tau\to \gamma \tau$.  The full path integral will therefore take the form of a sum over the modular group $SL(2,\ZZ)$, and the result is manifestly modular invariant.

There is one important subtlety, however, which is that certain elements of $SL(2,\ZZ)$ will act trivially on the geometry and hence not give distinct contributions to the path integral.  In the semi-classical limit these trivial elements of $SL(2,\ZZ)$ are easy to understand.
The transformation
 \be\label{g:bd}
T^n= \left({1~~n\atop 0~~1}\right) ~,\quad n\in \ZZ~,
 \ee
  shifts $\tau \to \tau + n$.  This does not change the topology of the resulting three-manifold, because adding a contractible cycle to a non-contractible cycle leaves the non-contractible cycle unchanged. 
Thus in the semi-classical regime, physically inequivalent saddles are labelled not by the group $SL(2,\ZZ)$ but by the coset $\Gamma=\Gamma_\infty\backslash SL(2,\ZZ)$ where $\Gamma_\infty$ is the group of shifts in equation \eqref{g:bd}. Elements of this coset are labelled by a pair of coprime integers $(c,d)$ with the restriction that $c\geq0$; these integers can be regarded as the bottom row of the $SL(2,\ZZ)$ matrix $\gamma = ({a ~ b \atop c ~ d})$.  The path integral \eqref{g:b}  then takes the form
\bea\label{g:bff}
Z_{\rm grav}(\tau,\bar\tau)&=&\sum_{\gamma\in \Gamma} Z_{\gamma}(\tau,\bar\tau)\cr
&=&\sum_{(c,d)=1\atop c\geq0} Z_{\rm vac}\left({a \tau + b\over c\tau + d},{a\bar \tau + b\over c\bar \tau + d}\right) \ .
\eea

In the strongly coupled regime, with central charge $c<1$, there is no good semiclassical approximation. However, as described above we still wish to organize the path integral as a sum over smooth geometries with fixed conformal structure at the boundary.  These geometries fall into distinct topological classes specified by the non-contractible cycle, so we can still formally write the partition function as
\be\label{g:bf}
Z_{\rm grav}=\sum_{\gamma \in \Gamma_c\backslash SL(2,\ZZ)} Z_{\rm vac}(\gamma\tau,\gamma \bar\tau)~.
\ee
As we will discover below there is an important difference between this formula and the semi-classical result \eqref{g:bff}.  The group $\Gamma_c$, which describes trivial gauge transformations, will take a different form when $c$ is less than one.  Whereas in the semi-classical case $\Gamma_c$ is simply the group of shifts \eqref{g:bd}, in the strongly coupled regime the identification of gauge inequivalent configurations will need to be modified.  We will see that there is an enhanced group of gauge symmetries, so that $\Gamma_c$ is a finite index subgroup of $SL(2,\ZZ)$.  This novel feature will render the sum easily computable.  

To summarize, we have made two fundamental assertions about the structure of the path integral of AdS${}_3$ quantum gravity.  First, given our boundary conditions we expect the partition function defined on a two-torus to be modular invariant, i.e.
\be\label{g:bfff}
Z_{\rm grav }(\tau,\bar\tau)=Z_{\rm grav }(\gamma \tau,\gamma\bar\tau)~.
\ee
Second, the path integral should be given by a sum over topologies \eqref{g:bf}. We now turn to the computation of $Z_{\rm vac}$, the contribution to the path integral from a fixed topology.%
   
\subsection{Fixed topology}

We have written the path integral as a sum over topologies, where each topology is labelled by an element of $SL(2,\ZZ)$.  The contribution to the path integral from a fixed topology should equal the classical action plus the set of quantum corrections  described in equation (\ref{g:b}).  As we are considering a strongly coupled theory there is no sense in which these quantum corrections are small.  It is not, for example, useful to organize them into a perturbative expansion in inverse powers of the coupling constant as would be natural in the semi-classical regime.  
We will need to obtain a full answer -- including all quantum effects -- in one fell swoop.  Fortunately, following the work of \cite{Castro:2011ui} (see also \cite{Castro:2010ce}) it is possible to do precisely this. 

We start by computing $Z_{\rm vac}(\tau,\bar\tau)$.  This is defined as the contribution to the path integral from metrics which are continuously connected to the thermal AdS${}_3$ solution \eqref{g:tads}.  This saddle is the Euclidean geometry obtained by analytically continuing the Lorentzian Anti-de Sitter metric to Euclidean signature and imposing the thermal identification  (\ref{g:baa}).  The partition function $Z_{\rm vac}$ therefore has an alternate interpretation, as the finite temperature partition function of  excitations around AdS${}_3$.  In particular, 
\be\label{g:tr}
Z_{\rm vac}(\tau,\bar\tau) = \Tr_{\rm vac} q^{L_0} {\bar q}^{\bar L_0}~~~~~~q=e^{2 \pi i \tau} \ ,
\ee
where the trace is over all states of the bulk theory which are smoothly connected to empty AdS${}_3$. 

In principle this trace could be computed by quantizing an appropriate classical phase space.  More precisely, we could consider the configuration space of all classical excitations of AdS${}_3$  which are continuously connected to the AdS ground state.  This configuration space is equal to the classical phase space of the theory, which comes equipped with a symplectic structure.  One could then hope to quantize this phase space explicitly in order to obtain the exact quantum Hilbert space and thus the trace \eqref{g:tr}.  

At first sight, one might think that this classical phase space is trivial because three dimensional general relativity has no local degrees of freedom.  Thus all classical solutions to Einstein's equations which are continuously connected to empty AdS are in fact diffeomorphic to AdS.  Thus they do not describe different geometries, but rather new metrics which are related to the original one by a change of coordinates.  So one might think that these new solutions are pure gauge and do not lead to new contributions to the partition function.
However, just as in the previous section, this is not quite the case: those diffeomorphisms which do not vanish sufficiently quickly at infinity lead to physical states and hence non-trivial contributions to the path integral.

To describe this space of states in more detail we must review the boundary conditions that are used to define our path integral.  The definition of AdS gravity requires a choice of fall-off conditions for the metric and matter fields. The standard Brown-Henneaux boundary conditions state that the metric is 
\be\label{g:ca}
{ds^2\over \ell^2} = d\rho^2 +\frac{1}{4}e^{2\rho} \left(- dt^2 + d\phi^2 \right) +O(\rho^0) ~,
\ee
at large $\rho$. We denote by $\zeta$ the vector which generates a diffeomorphism that preserves these boundary conditions, and by $H[\zeta]$ the corresponding phase space charge which generates this symmetry. 
The set of vectors $\zeta$ that preserve the boundary structure are 
\be\label{g:caa}
\zeta_n = e^{i n u}\left(\p_u - \half n^2 e^{-2 \rho} \p_v - i {n\over 2 } \p_\rho \right) + \cdots ,~~~~~{\bar \zeta_n} = e^{i n v}\left(\p_v -\half n^2 e^{-2 \rho} \p_u - i {n\over 2 } \p_\rho \right) + \cdots ~,
\ee
where  $u=\frac{1}{2}(t+\phi)$ and $v=\frac{1}{2}(t-\phi)$.  Here we have expanded in Fourier modes and use  ``$\cdots$'' to denote subleading corrections in $\rho$ that do not affect the charges $H[\zeta_n]$.  These subleading terms describe true gauge symmetries of the theory which do not lead to new physical states. 
The non-trivial symmetries are those for which $H[\zeta]$ is non-zero.  These give new physical states.

The remarkable observation of Brown and Henneaux is that the $H[\zeta_n]$ are computable and  satisfy the Virasoro algebra:
\be\label{g:cb}
i\{H[\zeta_n]\,,H[\zeta_m]\} = (n-m)H[\zeta_{n+m}]+{c\over 12} n(n^2-1)\delta_{n+m, 0}~,
\ee
and similarly for $H[\bar\zeta_n]$. The brackets in \eqref{g:cb} are Dirac brackets and  the central charge is %
\be
c={3\ell\over 2 G}~.
\ee

The states of the theory are obtained by acting on the ground state with one of these charges.  These states are therefore labelled by a choice of diffeomorphism $\zeta$, and are usually referred to as boundary gravitons. 
If we wish to compute the norm of a state $\zeta$, the standard expression  in terms of the symplectic structure (generalizing the scalar Klein-Gordon norm) reduces to the Dirac bracket of the Hamiltonian (see \cite{Castro:2011ui} for further details). In particular, the norm of the state $\zeta$ is
\be\label{g:cc}
||\zeta||^2 = \{  H[\zeta^*] \, , H[\zeta]  \}~.
\ee
The physical spectrum of the theory includes all states for which $|| \zeta||^2$ is positive.\footnote{For simplicity we assume unitarity in this subsection, but this is not required. In a non-unitary theory, a state is physical unless it has zero overlap with every other state, including itself.}
A state with zero norm describes an extra gauge symmetry, rather than a new physical excitation.

So far we have identified the classical phase space of the theory as the space of non-trivial diffeomorphisms applied to the vacuum.  At the level of finite, rather than infinitesimal, diffeomorphisms this space is (a central extension of) the infinite dimensional group ${\rm Diff}(S^1) \times {\rm Diff}(S^1)$.  Given this group structure one might hope that this  phase space could be quantized and $Z_{\rm vac}$ computed explicitly. When the central charge is greater than one this can be accomplished using the method of co-adjoint orbits \cite{Witten:1987ty}.  Unfortunately the quantization of the phase space in question appears to be much more difficult when $c<1$.\footnote{  
See however \cite{Aldaya:1989ra}, who discuss the quantization of this space using different methods and appear to obtain results consistent with ours below.}
We will therefore proceed indirectly, and argue that there is only one natural answer  consistent with the symmetries of the phase space in question.

To begin let us imagine canonically quantizing this theory by promoting the Dirac brackets in \eqref{g:cb} to commutators.  We will also re-label the generators as $L_n=H[\zeta_n]$ and $\bar L_n=H[\bar \zeta_n]$; these are now operators in the quantum theory.  
The vacuum $|0\rangle$ is the state annihilated by $L_0$, $\bar L_0$ as well as all of the corresponding Virasoro lowering operators.  This vacuum state is described semiclassically by empty AdS.    

The conformal symmetry of AdS${}_3$ gravity then strongly constrains the quantum spectrum of the theory.   
The operators $L_{n},\bar L_{n}$ with $n>0$ act as annihilation operators and hence lower the energy and momentum of a state. This implies that when acting on the vacuum  
\be
L_n|0\rangle =0 ~,\quad n>0~.
\ee
 Further, the vacuum is invariant under the rigid $SL(2,\RR)$ generators which generate the $SO(2,2)=SL(2,\RR)\times SL(2,\RR)$ isometries of empty AdS.  Thus
 \be\label{g:cd}
 L_{-1}|0\rangle = \bar{L}_{-1}|0\rangle=0~.
 \ee
Descendants of the vacuum are obtained by acting with raising operators $L_{-n},\bar L_{-n}$ with $n>1$. These states are the boundary gravitons, and generically take the form
 \be\label{g:cda}
 L_{-n_1}\cdots L_{-n_k}|0\rangle~,\quad n_i>1~.
 \ee
 It is important to note that these graviton states do not generate an ordinary Fock space and should not be regarded as free particles.  Instead, they should be thought of as Virasoro descendants of the vacuum, whose norm is computed via commutators as in (\ref{g:cc}).  This turns out to have important implications when $c<1$.

The partition function $Z_{\rm vac}$ is then the generating function which counts these states, according to equation \eqref{g:tr}.  
In computing $Z_{\rm vac}$ it is important to count only physical states with positive norm.
The result depends on whether $c$ is greater than or less than one.  When $c>1$, the norm of any boundary graviton \eqref{g:cda} is positive, with the exception of the null state $L_{-1}|0\rangle$.  This computation of the norm follows directly from the Virasoro algebra, using \eqref{g:cc}. The resulting trace is equal to the character of the full Verma module modded out by the null state, which is
\be\label{g:cf}
Z_{\rm vac} =\left|q^{(1-c)/24}{(1-q)\over \eta(\tau)} \right|^2 =\left|q^{-c/24}\left(1 + q^2 + q^3 + 2q^4 +2q^5 + 4q^6 + \cdots\right)\right|^2~.
\ee
Here $\eta(\tau)$ is the Dedekind eta function. The factor of $(1-q)$ accounts for the removal of $L_{-1}$ from the spectrum. 

When $c<1$ the result is more interesting.  The computation is in fact identical to the construction of irreducible representations of the Virasoro algebra using the Kac determinant formula.  This discussion is a standard part of the construction of minimal model conformal field theories with $c<1$.  We will therefore just state the results here and refer to the literature (e.g.\ \cite{DiFrancesco:1997nk}) for details. 

The first result is that when $c<1$ the theory contains negative norm states unless
\be
c=1-{6\over p(p+1)}~,
\ee
where $p>2$ is an integer. When $c$ takes one of these values there will still be null states with zero norm.  Therefore the partition function which counts physical (non-zero norm) states differs  from that of \eqref{g:cf}.  It is instead the character of a degenerate representation of the Virasoro algebra,
 \bea\label{g:ce}
 Z_{\rm vac}(\tau,\bar \tau)&=&{\rm Tr}_{\rm vac} \,q^{L_0}\bar q^{\bar L_0}= |\chi_{1,1}(\tau)|^2~,
 \eea
where\footnote{The notation $\chi_{1,1}$ is chosen to match minimal model conventions in the literature.} 
\be\label{g:cfa}
\chi_{1,1} = q^{(1-c)/24}{(1-q)\over \eta(\tau)} \left(1+\sum_{k=1}^\infty (-1)^k \left( q^{h_{1+k(p+1),(-1)^k +(1-(-1)^k)p/2}}+
 q^{h_{1,kp+(-1)^k +(1-(-1)^k)p/2}}\right)\right)~,
\ee
with
\be\label{hrs}
h_{r,s} = {(pr-(p+1)s)^2-1\over 4 p(p+1)}~.
\ee
The complicated power series multiplying $\eta(\tau)$ in \eqref{g:cfa} has the effect of removing the null states from the spectrum.  The result is that $\chi_{1,1}$ is the character of an irreducible representation of the Virasoro algebra.  We note that while we have focused on unitary theories, for which all states have positive norm,  similar expressions hold for the non-unitary minimal models.  In these cases the character $\chi_{1,1}$ includes states with both positive and negative norm.

The above analysis, although performed in  pure gravity, easily generalizes to other ``pure" theories of quantum gravity whose only perturbative states are the boundary excitations of empty AdS associated with an extended symmetry algebra. One generalization of our results is to pure supergravity \cite{Achucarro:1987vz,Achucarro:1989gm}.  Another is to the higher spin theories of \cite{Aragone:1983sz,Bergshoeff:1989ns}. In both cases the spectrum of boundary states is determined by an appropriate asymptotic symmetry algebra. These algebras are generalizations of the Virasoro algebra to include either fermionic generators and current algebras \cite{Banados:1998pi, Henneaux:1999ib} or (in the higher spin case) to a non-linear $\W_N$ algebra \cite{Henneaux:2010xg, Campoleoni:2010zq, Gaberdiel:2011wb, Campoleoni:2011hg}.

Let us now summarize our results.  When $c<1$ the partition function of AdS$_3$ quantum gravity is
\be
Z_{\rm grav}(\tau,\bar\tau)=\sum_{\gamma\in \Gamma_c \backslash SL(2,\ZZ) }Z_{\rm vac}(\gamma\tau, \gamma\bar \tau) \ .
\ee
Here $\Gamma_c$ is the subgroup of pure gauge symmetries, which will be described in more detail below.
The function $Z_{\rm vac}$ is equal to the vacuum character of a minimal model CFT
\be
Z_{\rm vac}(\tau,\bar \tau)= |\chi_{1,1}(\tau)|^2~,
\ee
where $\chi_{1,1}$ is given by \eqref{g:cfa} for $c<1$.  We emphasize that,  although our results are phrased in the language of minimal model CFTs, these results were derived directly in the bulk gravity theory.

\section{Pure Gravity and the Ising Model}\label{s:ising}

We now wish to compute explicitly the partition function of quantum gravity and show that it matches exactly that of a known, unitary conformal field theory.  In this section we will focus on the simplest case, that of Einstein gravity with $G=3 \ell$.  The partition function will equal that of the $c=\half$ minimal model, which describes the critical Ising model. In later sections we will  explore more complicated versions of the correspondence for other values of the central charge.  

\subsection{Review of the Ising model as a minimal model}

The simplest unitary minimal model has central charge 
\be
c = c(3,4) = \half \ .
\ee
Minimal models have a finite number of primary fields, which can be found by determining the degenerate 
irreducible representations of the Virasoro algebra.
When $c=\half$ there are three such degenerate, irreducible representations, with dimensions\footnote{Here $h$ denotes the scaling dimension of an operator on the plane.  It is related to the eigenvalue of $L_0$ on the cylinder by $L_0 = h-{c\over 24}$.}
\be\label{hs}
h_{1,1} = 0 \ , \quad h_{2,1} = \frac{1}{16}  \ , \quad h_{1,2} = \frac{1}{2} \ .
\ee
Further details about this minimal model, including explicit formulae for the characters of these representations, are given in appendix \ref{app:mm}.

The partition function of this theory can, by conformal symmetry, be written as a sum over the characters of these representations
\be
Z(\tau,\bar\tau)=\sum_{h,\bar h} N_{h,\bar h}\, \chi_h(\tau)\chi_{\bar h}(\bar \tau) \ .
\ee
Here $\chi_h$ are the characters of the irreducible Virasoro representations appearing in equation \eqref{hs} and  $\tau$ is the conformal structure of the torus.  The non-negative integers
$N_{h,\bar h}\in\mathbb{N}_0$ denote the multiplicities of the various representations.

Modular invariance
\be
Z(\tau) = Z(\tau+1) = Z(-1/\tau) ~,
\ee
imposes a strong constraint on the theory. 
A simple way  to impose invariance under $\tau\to\tau +1$ is to include only primaries with $h=\bar h$ in the spectrum; this follows from the transformations properties of $\chi_h$ which are spelled out in  \eqref{app:aa}--\eqref{app:ab}. Thus we are led to consider
the partition function
\be\label{im:ac}
Z_{\rm Ising}(\tau,\bar\tau) = | \chi_{1,1}(\tau) |^2 + |\chi_{2,1}(\tau)|^2 + |\chi_{1,2}(\tau)|^2 \ .
\ee
It is straightforward to check that \eqref{im:ac} is invariant under the action of $S$ as well. 

In fact, this minimal model has a simple interpretation in terms of the two dimensional Ising model.
The Ising model consists of a square lattice with spin variables at each site taking values $\pm 1$.  With a nearest neighbour interaction the theory has a conformally invariant critical point.  At this critical point the dynamics are described by the minimal model CFT with $c=\half$. 
The theory contains three local primary operators: the identity $1$, the spin operator $\sigma(z,\bar z)$, and the energy density $\varepsilon(z,\bar z)$, with conformal dimensions
\be
(h, \bar h)_1 = \left(0,0\right) ~~~\quad (h, \bar h)_{\varepsilon} = \left(\half, \half\right) ~~~\quad (h, \bar h)_{\sigma} = \left(\frac{1}{16}, \frac{1}{16}\right) \ .
\ee
These fields $1,\sigma,\varepsilon$ are 
obtained by multiplying together the left- and right-moving representations appearing in \eqref{hs}.
The partition function \eqref{im:ac} can then be seen to come from the states created by $1,\sigma$, and $\varepsilon$, respectively.  Remarkably, this conformal field theory can also be described in terms of a single free fermion.

\subsection{Duality}

We now compute the path integral of quantum gravity at $c=\half$ using the method described in section \ref{s:gravity} and show that it is equal to $Z_{\rm Ising}$.  
The contribution of thermal AdS to the path integral is simply the minimal model vacuum character 
\be
Z_{\rm vac}(\tau,\bar\tau) = |\chi_{1,1}(\tau)|^2 \ .
\ee
The full path integral is the sum over all modular images
\be\label{dogravsum}
Z_{\rm grav}(\tau,\bar\tau) = \sum_{\gamma\in \Gamma} |\chi_{1,1}(\gamma \tau)|^2 \ ,
\ee
where $\gamma \tau = \frac{a \tau + b}{c\tau + d}$.  Here $\Gamma=\Gamma_c \backslash SL(2,\ZZ)$ is the coset which labels physically inequivalent geometries.
We must now determine this coset.

The subgroup $\Gamma_c\subset SL(2,\ZZ)$ describes those diffeomorphisms which act trivially on the state of the system, and in particular leave the Hamiltonian and other observables unchanged. 
At first sight the identification of these residual gauge symmetries seems difficult, as we are working in the regime $c<1$ where we have no perturbative control over the bulk theory.  However, from the form of \eqref{dogravsum} is is clear that we should take $\Gamma_c$ to be the subgroup of $SL(2,\ZZ)$ which leaves the vacuum character $|\chi_{1,1}|^2$ invariant.  This $\Gamma_c$ is a finite index subgroup of $SL(2,\ZZ)$, so that the coset $\Gamma=\Gamma_c \backslash SL(2,\ZZ)$ is finite.  Thus it is necessary to include only a finite number of inequivalent topologies in the sum over geometries.  We will describe the general properties of this sum below, but before doing so let us work out the Ising model case in full detail.  

When $c=\half$ one can determine all of the inequivalent contributions to the partition function \eqref{dogravsum} explicitly.  In the basis $(\chi_{1,1} \ , \chi_{2,1} \ , \chi_{1,2})$ the modular matrices are given by equations (\ref{defmod}) and \eqref{app:its} in the appendix. 
Starting with the contribution from thermal AdS, $Z_{\rm vac}$, one can then apply $S$ and $T$ repeatedly to this `seed' contribution, obtaining new contributions to the partition function.  This process terminates after producing 24 inequivalent contributions, so $|\Gamma| = 24$ and the modular sum in (\ref{dogravsum}) is indeed finite.  Summing up these contributions gives
\be
Z_{\rm grav} = 8 Z_{\rm Ising} \ .
\ee
The overall factor of $8$ can be absorbed into the path integral measure. We conclude that the partition function of pure quantum gravity at $c=\half$ is equal to that of the Ising model.  Both the gravity path integral and the Ising model partition function are modular invariants constructed from the $\chi_{r,s}$,  so one might suspect that this equality is guaranteed by modular invariance.  This is indeed the case as will be discussed below.

We emphasize the key fact that the gravitational path integral with $c<1$ includes only a finite sum over topologies.  This differs from the infinite sum -- as in equation \eqref{g:bff} -- that arises in the $c>1$ regime.  The truncation to a finite number of terms was not imposed by hand.  It is required by the standard rules of path integration in a theory with gauge invariance.  In the quantum regime there is an enhanced gauge symmetry, rendering certain large gauge transformations trivial and collapsing the apparently infinite family of $SL(2,\ZZ)$ black holes to a finite number of inequivalent geometries.  In a sense this enhanced gauge invariance implies that many of the apparently distinct semi-classical black hole solutions of three dimensional gravity must disappear in this quantum regime.  This is clear if one attempts to interpret the primary states of the Ising model as black holes; there are only a finite number of such primaries, and they do not appear to have the large degeneracies associated with the semi-classical Bekenstein-Hawking formula.

\subsection{Tricritical Ising model}

The next conformal field theory on our list of unitary minimal models has central charge
\be\label{tim:aa}
c(4,5)={7\over 10}~. 
\ee
The theory contains primary fields with weights
\be\label{tim:ab}
h_{1,1}=0~,\quad h_{1,2}={7\over 16}~,\quad h_{1,3}={3\over 2}~,\quad h_{2,2}={3\over 80}~,\quad h_{2,3}={3\over 5}~,\quad h_{3,3}={1\over 10} ~.
\ee
As in the previous case, it is easy to construct a modular invariant partition function
\bea\label{tri}
Z_{\rm Tri-Ising}(\tau,\bar\tau)&=&\sum_{h,\bar h} N_{h,\bar h}\,\chi_h(\tau)\chi_{\bar h}(\bar \tau) 
=\sum_{r,s}|\chi_{r,s}(\tau)|^2~,
\eea
where $\chi_{r,s}$ are the Virasoro characters \eqref{app:cfa} for the fields listed in \eqref{tim:ab}. Further details of this minimal model are given in appendix \ref{app:mm}. 

This conformal field theory has a statistical interpretation as a simple generalization of the Ising model, where we now allow the spin variable to take values $\{0, \pm1\}$. This adds to the model a chemical potential associated to the fractional occupation number. This new parameter modifies the structure of the phase diagram, so that there is now a {\it tricritical} point where three phases meet: paramagnetic, ferromagnetic and a two-phase region.\footnote{The tricritical Ising model is secretly supersymmetric, but we will not exploit this feature here.}
This tricritical Ising model now has six primary operators, corresponding to the identity, three energy operators and two spin like operators; these give the six terms appearing in the partition function \eqref{tri}.

We now proceed to construct the gravitational dual of the tricritical Ising model. The discussion is very similar to that for the Ising model above. The gravitational path integral for $c={7\over 10}$ is
\be\label{tim:ba}
Z_{\rm grav}=\sum_{\gamma \in \Gamma} |\chi_{1,1}(\gamma\tau)|^2~,
\ee
where  $\chi_{1,1}$ is the vacuum contribution of thermal AdS, which equals the vacuum character of the tricritical Ising model. To find the subgroup $\Gamma=\Gamma_c \backslash SL(2,\ZZ)$ we must again identify those diffeomorphisms which leave the observables invariant. In practice, this means finding those elements of $SL(2,\ZZ)$ which leave $|\chi_{1,1}|$ invariant. 

The operator content of every $(p,p+1)$ minimal model is different.  Thus the representation of the modular matrices in the basis $\chi_{r,s}$  depends on $p$. This means that the set of pure gauge transformations $\Gamma_c$  is different for each value of the central charge.  For example, one can construct explicitly $\Gamma_c$ for the tricritical Ising model and show that it is an index 288 subgroup of $SL(2,\ZZ)$.  Thus there are 288 inequivalent contributions to the sum \eqref{tim:ba}.  This is in contrast to 24 for the critical Ising model.  In performing these explicit computations of the sum over geometries, the determination of $\Gamma_c$ is the most difficult part. 

For the tricritical Ising model, we find by explicitly computation
\be 
Z_{\rm grav} = \sum_{\gamma\in\Gamma_c\backslash SL(2,\ZZ)}|\chi_{1,1}(\gamma \tau)|^2=48 Z_{\rm Tri-Ising} \ .  
\ee
As before the overall coefficient can be absorbed into the path integral measure. This allows us to identify the partition function of the tricritical Ising with the path integral of pure quantum general relativity at $c={7\over10}$.

\section{Pure Gravity at Other Values of $c<1$}\label{s:pureother}

We now consider the general case, where $c$ takes any of the allowed minimal model values.  We will begin by investigating more thoroughly the implications of modular invariance, which lead to an improved understanding of the (tri)critical Ising model examples discussed above.

A minimal model conformal field theory -- and indeed any rational CFT -- has only finitely many primary fields.  The
corresponding characters\footnote{To simplify notation, we use the collective index $\lambda=(r,s)$ to label primary fields.} $\chi_\lambda$ will therefore form a finite-dimensional 
unitary representation of the modular group $SL(2,\ZZ)$.  That is to say, for every  $\gamma\in SL(2,\ZZ)$ we have 
\be
\chi_\lambda(\gamma \tau) = \sum_{\mu} M(\gamma)_{\lambda\mu}\, \chi_\mu(\tau) \ ,
\ee
where $M(\gamma)$ is a unitary  matrix.
 
For any of the allowed minimal model values of $c$ the partition function of quantum gravity will take the form
\be\label{gravZ}
Z_{\rm grav}  = \sum_{\gamma\in\Gamma}|\chi_{1,1}(\gamma \tau)|^2 \ .
\ee
Here, as above, the vacuum character $\chi_{1,1}$ is a ``seed" contribution from thermal AdS which is summed over modular images.  The coset $\Gamma$ is
\be
\Gamma = \Gamma_c  \backslash SL(2,\ZZ)  \ , \qquad
\Gamma_c = \left\{ \gamma \in SL(2,\ZZ) \ : \ |\chi_{1,1}(\gamma\tau)| = |\chi_{1,1}(\tau)| \right\} \ .
\ee
We note that $\Gamma_c$ always contains the parabolic subgroup of $SL(2,\ZZ)$ generated
by $T:\tau\mapsto \tau+1$.  More generally, $\Gamma_c$ contains the projective kernel of the modular representation $M(\gamma)$, i.e.\ the subgroup of $SL(2,\ZZ)$ which leaves all of the characters $\chi_\lambda$ invariant up to a phase. It was shown in
\cite{Bantay:2001ni} that, for every rational CFT, this is a finite index subgroup of $SL(2,\ZZ)$.  Hence
the sum \eqref{gravZ} involves only finitely many terms. This implies that there are no subtle issues involving the convergence of the sum over topologies.  This is in contrast with previous discussions of the sum over topologies in the black hole Farey tail.

The function $Z_{\rm grav}$ is by construction modular invariant.   Furthermore, given 
the structure of the modular transformations it must take the form
\be
Z_{\rm grav} = \sum_{\lambda,\mu} N_{\lambda\mu} \chi_\lambda\, \bar{\chi}_\mu 
\ee
for some constants $N_{\lambda\mu}$. In order for $Z_{\rm grav}$ to have an interpretation as 
a quantum mechanical partition function the constants $N_{\lambda\mu}$ must all, up to an overall scale factor, be non-negative integers with $N_{11}=1$ so that the vacuum is unique.  We will call a modular invariant combination of characters with these properties a {\em physical invariant}. 

In order to understand whether $Z_{\rm grav}$ will be a physical invariant or not, let us 
first consider the following more general problem. Suppose we start with some combination of 
characters of the form
\be\label{xis}
Z(X) = \sum_{\lambda,\mu} X_{\lambda \mu} \chi_\lambda\, \bar{\chi}_\mu  \ ,
\ee
where $X_{\lambda\mu}$ is a constant hermitian matrix. Under the modular transformation $\tau\mapsto \gamma\tau$,
$Z(X)$ becomes $Z(\gamma\cdot X)$, where
\be\label{modulaction}
\gamma\cdot X = M(\gamma) \, X \, M(\gamma)^\dagger \ .
\ee
The product on the right hand side is standard matrix multiplication.
We call $X$ {\em modular invariant} 
if $\gamma\cdot X=X$ for all $\gamma\in SL(2,\ZZ)$; then the corresponding combination of characters 
$Z(X)$ is a modular invariant function. Note that we do not assume that $Z(X)$ is a physical invariant in the sense that it satisfies any integrality or positivity properties. 

The space of matrices $X_{\lambda\mu}$ has a non-degenerate inner product
\be\label{inner}
(X,Y) = {\rm Tr}(X\, Y) 
\ee
that is clearly invariant under the action \eqref{modulaction} of the modular
group, 
\be
(X,Y) = (\gamma\cdot X,\gamma\cdot Y) \ .
\ee

Suppose now that $I_1,\ldots, I_k$ is a basis for the space of modular invariant hermitian matrices.
Without loss of  generality we may assume that the 
$I_j$ are orthonormal with respect to the inner product (\ref{inner}). Then, given
some seed $X$ the modular completion of $X$ will take the form
\be
\sum_{\gamma\in\Gamma} Z(\gamma\cdot X) = \sum_{j=1}^{k} c_j\,  Z(I_j) \ , 
\ee
where
\be
c_j =  \sum_{\gamma\in\Gamma}  (I_j,\gamma\cdot X) = \sum_{\gamma\in\Gamma} (\gamma^{-1} I_j, X)
       = |\Gamma|\, (I_j,X) \ .
\ee
Here we have used that $I_k$ is modular invariant, and $|\Gamma|$ equals the number of
elements in $\Gamma$. For a given rational CFT it is now (at least in principle) straightforward to
determine the modular completion of any given seed $X$: once we have classified all modular
invariants $I_j$, we only need to evaluate the inner products $(I_j,X)$ in 
order to compute the modular completion of $X$. 
The gravitational path integral \eqref{gravZ} can then be computed by choosing the seed $X_{\lambda\mu}=\delta_{\lambda,1}\, \delta_{\mu,1}$.

If the  CFT under consideration has more than one modular invariant combination of characters, then
the modular completion of any seed will lead to a linear combination of the different modular invariants
$Z(I_j)$.
This linear combination will typically not be a physical invariant. Thus we should expect that the 
construction will not lead to a physical modular invariant unless the CFT has only one
modular invariant combination of characters. 

On the other hand, if the CFT has a unique modular invariant, then this modular invariant must be 
physical since every rational CFT has at least one physical modular invariant given by the diagonal\footnote{The combination \eqref{diag} is modular invariant because the representation of the modular group is unitary.}
\be\label{diag}
Z(I_1) = \sum_{\lambda}\chi_\lambda\, \bar{\chi}_\lambda \ .
\ee
In this case the modular completion of {\em any} seed will lead to a physical invariant, since the modular completion
is modular invariant by construction.  Thus it will agree, up to an overall coefficient, with $Z(I_1)$. As 
we shall see momentarily, both the Ising and the tricritical Ising models fall into this class.

\subsection{Modular invariants of Virasoro minimal models}

For the Virasoro minimal models, the modular invariants have been classified by 
Cappelli, Itzykson \& Zuber  \cite{Cappelli:1986hf,Cappelli:1987xt}. Recall that the
minimal models are parametrised by two coprime positive integers $p<p'$, where 
\be\label{cdef}
c(p,p') = 1 - \frac{6 (p-p')^2}{ p p'} \ .
\ee
In the $(p,p')$-minimal model, the primary fields are labeled by $(r,s)$, where 
$r=1,\ldots,p'-1$ and $s=1,\ldots,p-1$ and we have the identification $(r,s)\sim (p'-r,p-s)$. 
Thus  the characters form a representation of $PSL(2,\ZZ)$ of dimension $(p-1)(p'-1)/2$. Explicit
formulae and further details of this representation are given in  appendix \ref{app:mm}. 

As reviewed in appendix \ref{app:ms}, the different 
modular invariant combinations of characters are labelled by pairs of integers $(d,d')$, where $d(d')$ is a divisor of $p(p')$.  The modular invariants can be described by matrices $X_{\lambda \mu}^{(d,d')}$ using the notation of equation \eqref{xis}, which satisfy 
\be\label{iden}
X^{(d,d')} = - X^{(\frac{p}{d},d')} = - X^{(d, \frac{p'}{d'})} = X^{(\frac{p}{d},\frac{p'}{d'})} \ .
\ee
The modular invariant $I_1$ exists for all Virasoro minimal models and corresponds to $d=d'=1$.  This is the diagonal invariant, also referred to as the `AA' invariant. 
For unitary minimal models, we have  $(p,p')=(m,m+1)$ with $m\geq 3$. For $m\geq 5$ there exists
at least one more modular invariant, namely $X^{(2,1)}$ when $m$ is even, or
$X^{(1,2)}$ when  $m$ is odd. 
The linear combination $X^{(1,1)}-X^{(1,2)}$ or  $X^{(1,1)}-X^{(2,1)}$ is physical and corresponds to 
the `AD' or the `DA' modular invariant, respectively.

In what follows we will illustrate the properties and consequences of this classification of modular invariants. We will start with a few simple examples, including the Ising and 
tricritical Ising models, before proceeding to the general case.  

\subsubsection{The Ising model $(p,p')=(3,4)$}

For the Ising model $(p,p')=(3,4)$, the possible divisors $(d,d')$ are 
$(1,1)$, $(1,2)$, $(1,4)$, $(3,1)$, $(3,2)$ and $(3,4)$.  The modular invariants with
$d=1,3$ and $d'=1,4$ are all (up to an overall sign) equal to the $I_1$ modular invariant.  The invariants with $(d,d')=(1,2)$ and $(d,d')=(3,2)$ vanish by equation (\ref{iden}). Thus the Ising
model has only one modular invariant. The argument around equation (\ref{diag}) then implies
that $Z_{\rm grav}$ is, up to a constant, equal to the $I_1$ modular invariant.

\subsubsection{The tricritical Ising model $(p,p')=(4,5)$}

The analysis for the tricritical Ising model is essentially identical. The possible divisors are 
$(1,1)$, $(1,5)$, $(2,1)$, $(2,5)$, $(4,1)$ and $(4,5)$.
Again, the modular invariants with $d=1,4$ and $d'=1,5$ are all (up to an overall sign) equal 
to the $I_1$ modular invariant, while the invariants $(2,1)$ and $(2,5)$ vanish by equation
(\ref{iden}). Thus the tricritical Ising model has also only one modular invariant, and again $Z_{\rm grav}$ is necessarily proportional to this modular invariant.

\subsubsection{The $(p,p')=(5,6)$ model}\label{sec:56}

Let us now consider the next unitary minimal model, with $(p,p')=(5,6)$. In this case
there are two different invariants, both of which are physical. In addition to the AA invariant 
with $(d,d')=(1,1)$ we also have a non-trivial AD invariant  $X^{(1,1)}-X^{(1,2)}$.  We should therefore not expect the $Z_{\rm grav}$ modular invariant, obtained as the modular completion of the seed $|\chi_{1,1}|^2$,  to be a physical invariant. Instead it will be a linear combination of the two physical modular invariants.

Let us denote by $X^{\rm AA}$ and $X^{\rm AD}$ the two canonically normalised physical modular invariants. 
An orthonormal basis for the modular invariants can be taken to be
\be
I_1 = \frac{1}{\sqrt{10}}\, X^{\rm AA} \ , \qquad
I_2 =  \sqrt{\frac{5}{48}}\, \Bigl(X^{\rm AD} - \frac{4}{5} X^{\rm AA} \Bigr) \ .
\ee
The inner products with the seed $|\chi_{1,1}|^2$ (i.e.\ with $X_{\lambda\mu}=\delta_{\lambda1}\delta_{\mu1}$) are
\be
c_1 = (I_1, X) = \frac{1}{\sqrt{10}} \ , \qquad 
c_2 = (I_2, X) = \frac{1}{\sqrt{240}} \ .
\ee
Thus, up to an overall numerical factor,
\begin{eqnarray}\label{56ex}
\sum_{\gamma\in \Gamma} |\chi_{1,1}(\gamma \tau)|^2 & \cong  & 
\frac{1}{\sqrt{10}} I_1 + \frac{1}{\sqrt{240}} I_2 
 \nonumber \\
& = & \frac{1}{12} \, X^{\rm AA} + \frac{1}{48}\, X^{\rm AD} 
\cong  \frac{4}{5} X^{\rm AA} + \frac{1}{5}\, X^{\rm AD}  \ .
\end{eqnarray}
In the last step we have rescaled
the partition function so that the vacuum representation appears with multiplicity one. Because
of the fractional coefficients the resulting invariant is not physical.\footnote{This final expression can also be checked by computing directly the sum over modular images.}  The coefficients appearing in the $q$-expansion are not positive integers, so we conclude that in this case the partition function of pure quantum gravity does not have a quantum mechanical interpretation.

\subsubsection{The general case}

In general the analysis of the $(p,p')=(m,m+1)=(5,6)$ model above appears to be representative of all unitary minimal
models with $m\geq 5$. Indeed, all of these models have at least a second (physical) modular
invariant.  Thus generically we expect fractional coefficients in the normalised modular
completion of the vacuum seed $|\chi_{1,1}|^2$ as in (\ref{56ex}). We have tested this for a number
of cases, and this conclusion seems to be fairly robust. 

We note, however, that if $p'\neq p+1$ -- so that the resulting models are non-unitary -- then one can find models with a unique modular invariant. In this case, the modular completion of the vacuum seed (or indeed any other seed) will necessarily give the physical modular 
invariant. The minimal models that have this property are characterized by the condition that 
$p$ and $p'$ are both prime or squares of primes; one can see from the 
discussion around (\ref{iden}) that there is a unique modular invariant in this case.

In general, however, it appears that the only cases where the path integral of general relativity equals that of a unitary CFT are the critical and tricritical Ising models.

\section{Higher Spin Gravity and the Potts Model}\label{s:higherspin}

So far we have limited our attention to pure gravity.  Now we will consider a wider class of bulk theories, and find candidate duals for several more exactly solvable CFTs.  The bulk theories that we consider are higher spin theories in AdS$_3$, which include higher spin gauge fields in addition to the graviton. Like pure gravity, higher spin gravity in three dimensions is locally trivial --- the only degrees of freedom are boundary excitations --- so the path integral can be computed exactly and compared with known CFTs.

The asymptotic symmetry algebra of higher spin gravity in AdS$_3$ is a $\W$-algebra \cite{Henneaux:2010xg,Campoleoni:2010zq}, which is an extension of the Virasoro algebra to include higher spin currents. This suggests that we look for dualities between quantum higher spin gravity and CFTs with extended conformal symmetry.  We will compute the path integral of higher spin gravity and find several examples where the resulting partition function matches that of a known CFT with extended  symmetry.  The simplest duality in this class relates the 3-state Potts model to $SL(3)$-gravity.  Both theories have $\W_3$ symmetry, and we will show that they have the same torus partition function.  This provides a higher spin analogue of the Ising model duality described above. Several other examples are also discussed below. 

\subsection{The Potts model and extended chiral algebras}\label{ss:extended}
We begin with a discussion of extended conformal symmetry in CFT, postponing the bulk interpretation until section \ref{ss:hsgrav}.  

In section \ref{sec:56}, we concluded that pure gravity does not give a quantum mechanically consistent partition function when $(p,p')=(5,6)$. At this value of the central charge, there are two possible minimal model conformal field theories, corresponding to the two modular invariants $X^{\rm AA}$ and $X^{\rm AD}$ described in section 4.1.3.   The gravitational partition function 
\be
\sum_{\gamma\in \Gamma} |\chi_{1,1}(\gamma \tau)|^2 \ 
\ee
does not give either of these partition functions.
However, this does not rule out a dual interpretation using a different theory of gravity.
The second of these minimal model conformal field theories -- the one with the non-diagonal `AD' type partition function -- is known as the 3-state Potts model.  An explicit formula for its partition function $Z_{\rm potts}$ is given in equation \eqref{inv2}.  It turns out that $Z_{\rm potts}$ can be written as a sum over modular images starting from a different vacuum seed.  For the Potts model, we observe that
\be\label{pottsum}
\sum_{\gamma \in \Gamma}|\chi_{1,1}(\gamma \tau) + \chi_{1,4}(\gamma \tau)|^2 = 12 \,Z_{\rm potts} \ ,
\ee
where the sum is over the 48 distinct images of the vacuum seed under $SL(2,\ZZ)$. Here $\chi_{1,4}$ is the character built on a Virasoro primary with dimension 3.  

The expression (\ref{pottsum}) has a natural CFT interpretation.  The conserved spin-3 current of the Potts model extends the chiral algebra of the theory from the Virasoro algebra to $\W_3$.  Thus states can be organized into representations of $\W_3$.  The vacuum representation of this algebra now contains not just Virasoro descendants but also spin-3 descendants, so
\be
\chi_{\rm vac}^{(3)}(\tau) = \chi_{1,1}(\tau) + \chi_{1,4}(\tau)~,
\ee
where $\chi^{(3)}$ denotes a $\W_3$ character at $c=\frac{4}{5}$.  Therefore (\ref{pottsum}) can be written
\be\label{pottsw}
\sum_{\gamma \in \Gamma}|\chi_{\rm vac}^{(3)}(\gamma\tau)|^2 = 12 \,Z_{\rm potts} \ .
\ee
This is very similar to the identity for the Ising model, except that the vacuum contribution is a $\W_3$ character rather than a Virasoro character.  The equality (\ref{pottsw}) is guaranteed by modular invariance: there is only one modular invariant combination of $\W_3$ characters at $c=\frac{4}{5}$ 
(see e.g.\ \cite{Gannon:1994km}). We note that for the purposes of computing the torus partition function we only care about the functional dependence of this partition function on $\tau$.  However, this masks some of the structure of the theory, as there are 
conjugate representations of $\W_3$ which happen to have the same character.  Thus the partition function can actually be written in terms of $\W_3$ characters in many different ways. As we will discuss in more detail in section~\ref{sec:para} this will have implications for the computation of higher genus partition functions, although it is invisible in the torus partition function.

Other $(p,p+1)$ Virasoro minimal models in the $D$-series similarly have extended chiral algebras when $p\equiv1,2 (\rm{mod\ } 4)$. The next case, $p=6$ and $c=\frac{6}{7}$, is the tricritical Potts model.  This CFT has a conserved dimension-5 current in addition to the stress tensor at dimension 2, so the chiral algebra is denoted $\W(2,5)$.  The $\W(2,5)$ vacuum character at $c=\frac{6}{7}$ is 
\be
\chi_{\rm vac}^{(5)} = \chi_{1,1} + \chi_{1,5}~,
\ee
where $\chi_{1,5}$ contains the dimension-5 Virasoro primary.  Summing over modular images, we find
\be\label{esum}
\sum_{\gamma \in \Gamma}|\chi_{\rm vac}^{(5)}(\gamma\tau)|^2 = 16 \, Z_{\rm tri-potts} \ .
\ee

For $p=9,10,13$, with chiral algebras $\W(2,14)$, $\W(2,18)$, and $\W(2,33)$ respectively, the modular sum reproduces the CFT partition function. However, this pattern does not continue indefinitely. For $p=14$, it fails: the modular sum produces a partition function with some negative integer coefficients, which is not the partition function of a unitary CFT.  

As discussed in Appendix \ref{app:twoinvariants}, this can be understood from the Cappelli-Itzykson-Zuber classification of modular invariants.  For $p>4$, all minimal models have an AA invariant and an AD or DA invariant (and some have an exceptional E invariant).  These are the only physical invariants -- i.e.\ the only ones with positive integer coefficients -- but in general there are also unphysical invariants.  The unphysical invariants first appear at $p=14$, and account for the negative integers in the gravity answer at this level. Conditions for a unitary model to have only physical invariants are given in Appendix \ref{app:twoinvariants}; in these cases the sum over modular images of the extended vacuum character reproduces the AD or DA-type CFT partition function.  Otherwise we expect a contribution from unphysical invariants, which is confirmed in detail for $p=14$ in Appendix \ref{app:unphysical}.
These facts are summarized in the first three columns of Table \ref{table:results} in section 6.

\subsection{Extended chiral algebras from higher spin gravity}\label{ss:hsgrav}

We have identified an infinite class of unitary CFTs --- Virasoro models with $p=5,6,9,10,13$, plus those in Appendix \ref{app:twoinvariants} --- with extended chiral algebras, where the modular sum of the extended vacuum character produces a consistent CFT partition function.  Now we attempt to construct bulk theories whose contribution around thermal AdS is given by the extended vacuum character of these CFTs.  We will succeed only for $p=5,6$; the results are summarized in the final column of Table \ref{table:results}.

As discussed above, the natural candidate dual for a theory with extended conformal symmetry is a higher spin theory.  There are many different theories of higher spin gravity in AdS$_3$, specified by a Lie algebra $g$ and an embedding $\rho$ of  $sl(2) \subseteq g$.  The action is the sum of two Chern-Simons actions,
\be\label{csaction}
S = k_{\rm cs} I_{\rm CS}[A] - k_{\rm cs} I_{\rm CS}[\bar{A}] \ , \quad \quad 
I_{\rm CS}[A] = \frac{1}{4\pi}\int \left(A dA + \frac{2}{3}A^3\right)~,
\ee
where the gauge fields $A,\bar{A}$ take values in $g$.\footnote{The subscript on $k_{\rm cs}$ is to distinguish the level of the Chern-Simons theory from the level $k$ in the coset construction of minimal models. In Lorentzian signature, $A$ and $\bar{A}$ are independent, each taking values in the split real form of $g$. In Euclidean signature, $A$ takes values in the complex Lie algebra $g$ and $\bar{A}$ is its complex conjugate.} Pure gravity is the case $g=sl(2)$, studied in the Chern-Simons language by Achucarro and Townsend \cite{Achucarro:1987vz} and Witten \cite{Witten:1988hc}. Blencowe \cite{Blencowe:1988gj} (see also \cite{Bergshoeff:1989ns,Vasiliev:1989qh}) generalized this construction to the infinite-rank higher spin algebra $g = \rm{shs}(1,2)$ of Fradkin and Vasiliev \cite{Fradkin:1986ka,Vasiliev:1986qx}.  The choice $g = sl(N)$ was studied recently in \cite{Campoleoni:2010zq}.

The choice of embedding $sl(2) \subseteq g$ defines the metric components of the gauge field.  This determines the boundary conditions and the spectrum; for example choosing $g = sl(N,\RR)$ with the principal embedding of $sl(2,\RR)$, the adjoint of $sl(N,\RR)$ decomposes as
\be
\mbox{adj} = \bigoplus_{r=1}^{N-1}D_r \ ,
\ee
where $D_r$ is the spin-$r$ representation of $sl(2,\RR)$.  We see that these representations under $sl(2,\RR)$ correspond to a tower of higher spin fields with spins $s=2,3,\dots,N$.\footnote{In four dimensions the situation is different: all known theories have an infinite tower of higher spin fields \cite{Vasiliev:2001ur}.}

Let us consider $SL(3)$ higher spin gravity with the principal embedding, with the coupling constant chosen so that $c = \frac{4}{5}$.  Since this theory has $\W_3$ asymptotic symmetry \cite{Campoleoni:2011hg} it is natural to ask whether its path integral gives the partition function of the Potts model.  This is indeed the case.  The contribution to the path integral around thermal AdS can be computed much like in pure gravity; the only degrees of freedom are boundary gravitons and their higher spin cousins.  Generalizing the argument of section \ref{s:pureother} to include higher spin boundary excitations, we find for $SL(3)$ gravity
\be
Z_{\rm vac} = |\chi^{(3)}_{\rm vac}(\tau)|^2 \ ,
\ee
where $\chi_{\rm vac}^{(3)}$ is the $\W_3$ character defined in section \ref{ss:extended}.
The sum over saddlepoints is the sum over modular images (\ref{pottsw}), so we find
\be
Z_{\rm SL(3)\ gravity}^{c=4/5}  = 12\, Z_{\rm potts} \ .
\ee
Note that we have not included any new saddle points in the path integral; the only known classical solutions of higher spin gravity that obey the boundary conditions specified in section \ref{s:gravity} are those of ordinary gravity -- the $SL(2,\ZZ)$ family of black holes.  Higher spin black holes which are inequivalent to BTZ have been constructed in \cite{Gutperle:2011kf,Ammon:2011nk, Castro:2011fm}, but they do not obey asymptotically AdS boundary conditions so do not contribute to our path integral.

For a general theory of higher spin gravity based on the Lie algebra $g$ with a choice of embedding $\rho$, the asymptotic symmetry algebra is given by Drinfeld-Sokolov (DS) reduction.  DS reduction \cite{Drinfeld:1984qv,Fateev:1987zh}, reviewed in \cite{Bouwknegt:1992wg}, is an algebraic construction that associates a $\W$-algebra to any affine algebra $\hat{g}_{k_{\rm cs}}$ and $sl(2)$ embedding $\rho$.  This algebra, $\W_{DS}(g,k_{\rm cs},\rho)$, is defined by imposing a constraint on the affine algebra $\hat{g}_{k_{\rm cs}}$.  The relation between DS reduction and the asymptotic symmetries of higher spin gravity was pointed out in \cite{Henneaux:2010xg,Campoleoni:2010zq} and further explored in \cite{Gaberdiel:2011wb,Campoleoni:2011hg,Gaberdiel:2011zw}.  The essential point is as follows.  The conserved charges of Chern-Simons gauge theory generate an affine algebra $\hat{g}_{k_{\rm cs}}$.  Higher spin gravity has the same action (\ref{csaction}) as Chern-Simons gauge theory, but is not quite identical because (among other reasons \cite{Witten:2007kt}) boundary conditions must be imposed on the metric and higher spin fields.  The AdS$_3$ boundary conditions turn out to be exactly the constraints on $\hat{g}_{k_{\rm cs}}$ that implement the DS reduction.  For pure gravity, this reduces  the $\hat{sl}(2)_{k_{\rm cs}}$ affine symmetry of Chern-Simons theory to the Virasoro symmetry found by Brown and Henneaux \cite{Verlinde:1989ua,Coussaert:1995zp}. For higher spin gravity based on $g,\rho$ at level $k_{\rm cs}$, the boundary conditions reduce the asymptotic symmetries to $\W_{DS}(g,k_{\rm cs},\rho)$.

This suggests a candidate dual for the tricritical Potts model with chiral algebra $\W(2,5)$.  The $\W(2,5)$ algebra can be obtained by DS reduction of $E_6$ at central charge $c = \frac{6}{7}$ \cite{Bouwknegt:1988sv}.  Therefore the asymptotic symmetry algebra of $E_6$ higher spin gravity, though generically much larger, truncates to $\W(2,5)$ at this special value of the coupling.  The excitations around thermal AdS produce the vacuum character of $\W(2,5)$, and thus according to (\ref{esum}) we have
\be
Z_{E_6\ \rm{gravity}}^{c=6/7} = 16 \ Z_{\rm tri-potts} \ .
\ee

The next case to consider is $p=9$, with chiral algebra $\W(2,14)$.  We know of no way to construct this algebra by DS reduction, so cannot conjecture any bulk dual with the appropriate partition function around thermal AdS.  
Similarly for $p=10,13$ and the other cases in Appendix \ref{app:twoinvariants}.  Although the CFT partition function can be written as a sum over modular images, we know of no bulk theory whose path integral produces this partition function.

\subsection{Parafermions}\label{sec:para}

One can also construct minimal models associated to the extended chiral algebra $\W_N$. Here $\W_N$ denotes $\W(2,3,\dots,N)$, the algebra with higher spin conserved currents of all spins up to $N$ in addition to the stress tensor at dimension 2.  In this section we ask whether these CFTs may be dual to pure higher spin gravity. 

The $\W_N$ minimal model CFTs, reviewed in \cite{Bouwknegt:1992wg}, can be obtained as cosets
\be\label{wcoset}
\frac{\hat{su}(N)_k \times \hat{su}(N)_1}{\hat{su}(N)_{k+1}} \ ,
\ee
from which their characters, modular $S$ and $T$ matrices, and partition functions can be computed exactly. $\W_2$ is the Virasoro algebra, so setting $N=2$ we recover the unitary Virasoro minimal models where $p=k+2$.  In general, the central charge of a $\W_N$ minimal model is
\be\label{wnc}
c = (N-1)\left(1 - \frac{N(N+1)}{(N+k)(N+k+1)}\right) \ .
\ee
Setting $N=3,k=1$ gives $c=\frac{4}{5}$, the central charge of the Potts model. Thus in addition to being the $(5,6)$ Virasoro minimal model, it is the first $\W_3$ minimal model.  Although this first example is the Potts model, generically these are new CFTs which do not appear among the Virasoro models.  Since $SL(N)$ gravity has asymptotic symmetry algebra $\W_N$ \cite{Campoleoni:2010zq}, it is natural to ask when the path integral of $SL(N)$ gravity produces a consistent partition function for one of the $\W_N$ minimal models.   We expect that this will only be the case when the characters of the chiral algebra have a unique modular invariant combination.  

Some examples are provided by the $\W_N$ minimal models at level $k=1$, known as parafermion theories, with central charge
\be
c = 2\frac{N-1}{N+2} \ .
\ee  
$N=2$ is the Ising model and $N=3$ is the Potts model.  $N=4$ parafermions have chiral algebra $\W(1,2,3,4)$, which is bigger than $\W_N$ due to the appearance of the spin-1 current.  The sum over modular images of the vacuum character of this algebra produces a consistent partition function, but we do not know of any algebra whose DS reduction gives $\W(1,2,3,4)$, so there is no candidate bulk dual.

	Parafermions with $N=5,6,7,8$ do not have any extra conserved currents --- the chiral algebra is simply $\W_N$ in each case --- so the natural guess for a bulk dual is $SL(N)$ gravity.  The thermal AdS contribution to the bulk path integral of $SL(N)$ gravity is the vacuum character of $\W_N$.  To compute the sum over modular images, we need the $S$ and $T$ matrix of the coset model (see for example \cite{DiFrancesco:1997nk,Gannon:2001py}).  $T$ is the obvious generalization of (\ref{app:aa}), and $S$ is the product of $S$-matrices for the affine algebras appearing in (\ref{wcoset}).  Performing the sum over inequivalent saddlepoints, we find that the bulk path integral gives a consistent CFT partition function for all of these cases.  It would be interesting to see whether this continues at larger $N$.

Although the CFT partition functions obtained in this way are consistent, they have a peculiar feature that is illustrated by the 3-state Potts model.  Whether we view this theory as the $k=2$ Virasoro model or the $k=1$ $\W_3$ model, it is the same CFT, but the $\W_3$ partition function contains in a sense more information.  In $\W_3$ language, the Potts model consists of $\W_3$ primaries labelled by representations of $SU(3)$,
\newcommand{\ff}{\mbox{f}}
\be\label{w3spec}
\phi_1 = (0; 0) , \quad 
\phi_2 = (\ff; 0) , \quad \phi_2^c =(\bar{\ff}; 0), \quad
\phi_3 = (0; \ff) , \quad \phi_3^c = (0; \bar{\ff}), \quad
\phi_4 = (\ff; \bar{\ff}) \ ,
\ee
where $\ff$ is the fundamental and $\bar{\ff}$ is the antifundamental.  The corresponding conformal weights are $h_{1,2,3,4} = 0, \frac{2}{3}, \frac{1}{15}, \frac{2}{5}$, and $\phi^c$ denotes a conjugate field, which has the same conformal weight but opposite eigenvalue under the spin-3 current.  Thus the partition function is encoded in a $6\times 6$ matrix of positive integers,
\be\label{secondmod}
Z_{\rm potts}(\tau,\bar\tau) = \sum_{i,j=1}^{6}R_{ij} \, S_i(\tau) \overline{S_j(\tau)} \ ,
\ee
where $S_i$ denote the $\W_3$ characters for (\ref{w3spec}). 

Viewed as a function of $\tau$, the partition function does not completely fix the matrix $R_{ij}$ because conjugate representations have the same character,
\be
S_2 = S_2^c \ , \quad S_3 = S_3^c \ .
\ee
Therefore if we only know $Z$ as a function of $\tau$, we cannot distinguish between the diagonal spectrum 
\be
Z = |S_1|^2 + |S_2|^2 + |S_2^c|^2 + |S_3|^2 + |S_3^c|^2 + |S_4|^2\ ,
\ee
and the mixed invariant
\be\label{zmixed}
Z = |S_1|^2 + \half|S_2 + S_2^c|^2 +\half|S_3 + S_3^c|^2 + |S_4|^2  \ .
\ee
The first is the spectrum of the Potts model, while the second is inconsistent since it has half-integer coefficients when the vacuum  is unit normalized.  These are identical functions of $\tau$ which describe different operator content.

When the partition function of $SL(3)$ higher spin gravity is computed by the prescription in section \ref{s:pureother}, it gives the matrix $R_{ij}$ in (\ref{secondmod}) corresponding to the latter spectrum, not the diagonal.  This  does not by itself imply any inconsistency, because the gravity path integral is only a function of $\tau$; it cannot distinguish between the diagonal and mixed invariants.  Still, it leaves open the possibility that $SL(3)$ gravity is \textit{not} the 3-state Potts model, but rather an inconsistent theory with the spectrum (\ref{zmixed}).  The two possibilities cannot be distinguished at genus one, but would lead to different correlation functions, and different partition functions at genus two or higher.

\section{Discussion}\label{s:discussion}

We have provided evidence that several theories of quantum gravity are dual to simple, exactly solvable minimal conformal field theories.  Table \ref{table:results} summarizes our results.
\begin{table}
\begin{tabular*}{\textwidth}{c|c|c|c}
$c$ & \hspace{2cm} Minimal model CFT \hspace{2cm}  &\ \  Algebra \ \  &  Bulk dual \\ \hline\hline
$\frac{1}{2}$ & $N=2, k=1$ (Ising) & Virasoro  & Pure gravity\\
$\frac{7}{10}$ & $N=2, k=2$ (Tricritical Ising) & Virasoro  & Pure gravity\\
$\frac{4}{5}$ & $N=2,k=3$ or $N=3,k=1$ (Potts) & $\W_3$ & $SL(3)$ gravity\\
$\frac{6}{7}$ & $N=2,k=4$ (Tricritical Potts) & $\W(2,5)$ & $E_6$ gravity\\
$ 2\frac{N-1}{N+2}$ & $N=5,6,7,8$, $k=1$ (Parafermions) &  $\W_N$  & $SL(N)$ gravity\\
 1 & $N=4,k=1$ (Parafermions) & $\W(1,2,3,4)$  & unknown \\
 & $N=2,k=7,8,11,\dots$, see Appendix \ref{app:twoinvariants} &  & unknown
\end{tabular*}
\caption{\label{table:results}\small Unitary CFTs for which the torus partition function has a natural interpretation as a sum over geometries. Where the bulk dual is `unknown', there is no known way to produce the correct chiral algebra by Drinfeld-Sokolov reduction, as discussed in section \ref{ss:hsgrav}. The last entry is an infinite class of models with central charge (\ref{wnc}), and extended chiral algebras as discussed in section \ref{ss:extended} and Appendix \ref{app:twoinvariants}.}
\end{table}
We emphasize that this is the first time that the partition function of general relativity has been computed in three dimensions and found to give a result which describes a unitary quantum theory.

At this point these dualities should be regarded as conjectural.  The torus partition function of a CFT includes all data about its spectrum, but no non-trivial information about correlation functions. So there is much more that could be checked.
The data in these correlation functions is encoded in the higher genus CFT partition functions, which can in principle be matched with an analogous gravity path integral computation at higher genus.  This computation is more complicated than the one described in the present paper, but may be possible.\footnote{A computation of this sort was described in \cite{Yin:2007gv, Yin:2007at}, under the assumption that the partition function of pure gravity factorizes holomorphically.  This makes the computation technically easier, but cannot be justified from the bulk gravity point of view unless one includes complex saddle points or considers a chiral theory (such as chiral gravity).}  In the case of the Ising model  the higher genus partition functions are well studied so there is a possibility of a precise check.  One might hope to prove that the all-genus partition function of Einstein gravity at $c=1/2$ agrees with the Ising model, which would  practically amount to a proof of the duality. 

There is a second natural question to ask at this point, which  is why only certain minimal model conformal field theories appear in Table \ref{table:results}.  Of all possible minimal models, what is special about these theories which allows them to be interpreted as pure theories of quantum gravity?  Although we cannot answer this question definitively, we can make one suggestive observation.  Let us consider those theories which are dual to pure gravity, the critical and tricritical Ising models.  These two theories have the property that all of the non-trivial states in the theory obey the bound
\be\label{eq:magic}
h>{c\over 24} \ .
\ee
Moreover, among all the Virasoro minimal models with diagonal modular invariant these are the only two theories with this property!  All other minimal models admit primary states below the bound (\ref{eq:magic}).

From the bulk gravity point of view, this bound is completely natural.  
Three dimensional AdS gravity possesses BTZ black holes which are separated from the AdS ground state by a gap.  These black holes correspond to states in the dual CFT with weight larger than $c/24$, i.e.\ they lie in the ``Cardy regime" of the dual CFT. Indeed, this is why it is possible to describe the BTZ entropy using the Cardy formula of the dual CFT \cite{Strominger:1997eq}.  Thus the critical and tricritical Ising models are the only two diagonal minimal model CFTs with the property that all primary states can be interpreted as black holes.  
What is remarkable is that the bound $h>c/24$, which was justified using semiclassical gravitational reasoning,  appears to apply in the strongly coupled regime.  It would be interesting to understand the extent to which the primary operators of these CFTs can be interpreted as black holes.  These are highly quantum theories of gravity where usual semi-classical notions fail to apply, but there may still be some useful sense in which these states share the properties of black holes.

All other minimal models have primaries which may be interpreted as matter fields, and hence may be dual not to pure gravity but rather to something more complicated.  One might hope to generalize the analysis of the modular sums in sections 4 and 5 to construct other 
candidate vacuum partition functions whose sum over the modular group gives a minimal model partition function.  However, it is not clear that this is possible.  We have not been able to find a natural seed that 
would lead to a physical modular invariant even for a simple example (see appendix~\ref{app:unphysical}).

Finally, we note that the results described in Table \ref{table:results} are not exhaustive.  
For example, we have restricted our attention to models without supersymmetry, but we expect that the generalization of this to supersymmetric models would be straightforward.  It would be interesting to explore this and other classes of examples in detail.  It may even be possible to find a class of unitary examples which have a large central charge limit; this would then provide simple examples of theories of quantum gravity with a semi-classical limit.

\section*{Acknowledgements}

We thank M. Cheng, T. Gannon, H. Ooguri, E. Verlinde and X. Yin for useful discussions and comments. The work of 
MRG and RV was partially supported by the Swiss National Science Foundation. 
The work of TH is supported in part by U.S. Department of Energy grant DE-FG02-90ER40542.  The work 
of AC and AM is supported in part by the National Science and Engineering Research Council of Canada.  This material is based upon work supported in part by the National Science Foundation under Grant No. 1066293 and the hospitality of the Aspen Center for Physics.  The work of AC was supported in part by the National Science Foundation under Grant No. NSF PHY05-51164 and the hospitality of the Kavli Institute for Theoretical  Physics, UCSB. 

\appendix

\section{Minimal Models}\label{app:mm}

\subsection{Overview}

The Virasoro minimal models are labeled by coprime integers $(p,p')$ with $p'>p>2$ and central charge given by (\ref{virc}). The degenerate representations of the Virasoro algebra at $c= c(p,p') < 1$ are labeled by integers $(r,s)$ with
\be\label{app:rs}
1 \leq r \leq p'-1\ , \quad 1 \leq s \leq p-1 \ ,   \quad (r,s) \sim (p'-r,p-s) \ . 
\ee
The conformal dimensions of these representations are
\be
h_{r,s} = \frac{(p r - p' s)^2 - (p-p')^2}{4 pp'} \ ,
\ee
and their characters are denoted by
\be\label{app:ac}
\chi_{r,s}(\tau) = \tr \, q^{L_0}  \ , \quad q\equiv e^{2\pi i \tau} \ ,
\ee
with $L_0 = h- c/24$. These characters transform into one another under modular transformations
\be
T: \tau \to \tau + 1 \ , \quad S: \tau \to -1/\tau \ .
\ee
Organizing the characters into a vector $\chi_{\mu}$ where $\mu = (r,s)$, the transformation rules are
\be\label{defmod}
\chi_\mu(\tau+1) = \sum_\nu T_{\mu \nu}\chi_{\nu}(\tau) \ , \quad 
\chi_\mu(-1/\tau) = \sum_\nu S_{\mu \nu}\chi_\nu(\tau) \ ,
\ee
where
\bea\label{app:aa}
T_{rs;\rho\sigma}=\delta_{r,\rho}\delta_{s,\sigma}\exp\bigl[2\pi i (h_{r,s}-\tfrac{c}{24})\bigr]~
\eea
and
\bea\label{app:ab}
S_{rs;\rho\sigma}=2\sqrt{2\over pp'}(-1)^{1+s\rho+r\sigma}
\sin\left(\pi {p\over p'}r\rho\right)\sin\left(\pi {p'\over p}s\sigma\right)~.
\eea
To write explicit expressions for the minimal model characters we will use, given the equivalence relation (\ref{app:rs}), the values of $(r, s)$ such that the product $rs$ is minimized.  With this choice
 \be\label{app:cfa}
\chi_{r,s}(\tau) = K_{r,s}^{(p,p')}(\tau)-K_{r,-s}^{(p,p')}(\tau)
\ee
where
\be
K_{r,s}^{(p,p')}(\tau)={1\over \eta(\tau)}\sum_{n\in \ZZ}q^{(2pp'n+pr-p's)^2/4pp'} .
\ee

\subsection{ADE classification}

We now review the construction of modular invariant partition functions.  For a given chiral algebra we wish to construct all 
partition functions
\be\label{zrcft}
Z(\tau,\bar\tau)=\sum_{h,\bar h} N_{h,\bar h}\, \chi_h(\tau)\chi_{\bar h}(\bar \tau)~,
\ee
that are  modular invariant and have a unique vacuum. These  conditions imply that the matrix  
$N_{h,\bar h}$  satisfies
\bea\label{condmi}
N_{h,\bar h}\in \NN_0~,\quad N_{0,\bar 0}=1~,\quad N T = T N~,\quad  N S = S N~.
\eea
This classification problem  for rational CFTs is a well posed, but difficult, algebraic problem. For rational theories based on the $sl(2)$ algebra there is a one-to-one correspondence between modular invariants and pairs of simply laced Lie algebras with Coxeter numbers $p'$ and $p$. This is the ADE classification of minimal models developed in  \cite{Cappelli:1986hf,Cappelli:1987xt,Kato:1987td}. For a quick version of the construction see \cite{Gannon:1999, DiFrancesco:1997nk}.

 We do not intend to review the entire ADE classification. For the purpose of our discussion it will be sufficient to highlight two classes of physical modular invariants: diagonal (AA) and block diagonal (AD, DA). 
 
\subsubsection*{Diagonal Invariants} 
Invariance under the action of $T$ restricts the relative values of the left and right weights in \eqref{zrcft} to 
\be\label{tinv}
h-\bar h=0\mod 1~.
\ee
One natural solution is to have non-zero entries in $N$ only for  $h=\bar h$. Together with the remaining conditions   \eqref{condmi} the only solution is the diagonal modular invariant
\be
Z_{\rm AA}=\sum_{r,s}|\chi_{r,s}|^2~.
\ee
Modular invariance is guaranteed because $S$ is a unitary matrix. The subscripts AA refer to the $A_n$ algebra of the ADE classification.
 
\subsubsection*{Block diagonal Invariants}
The next to simplest solution to  \eqref{tinv} involves linear combinations of characters whose conformal dimensions differ by integers (so that the whole linear combination is invariant under the $T$-transformation). Furthermore, one needs to require that the $S$-transformation
relates the relevant linear combinations to one another. Then one can consider the usual diagonal
modular invariant of this `extended theory'. 

 For example, for $(p,p')$ with $p'=4m+2$ and $m>1$, the   DA modular invariant is of this type
 \be
 Z_{\rm DA}= {1\over 2}\sum_{s=1}^{p-1}\left[\sum_{r \,{\rm odd}=1}^{2m-1}|\chi_{r,s}+\chi_{4m+2-r,s}|^2+2|\chi_{2m+1,s}|^2\right]~.
 \ee
 A similar formula (with the roles of $r$ and $s$ interchanged) applies for the 
AD invariant with $p=4m+2$.

\subsection{Examples}\label{app:examples}

\subsubsection*{Ising Model}

The Ising model is the $(3,4)$ Virasoro minimal model  with central charge $c=\half$. The characters $\chi_{1,1}$, $\chi_{1,2}$, $\chi_{2,1}$ correspond to fields of weight $h=0,\frac{1}{2},\frac{1}{16}$, respectively. 
These characters transform in the three dimensional representation of $SL_2(\ZZ)$ with
\be\label{app:its}
T=\begin{pmatrix}e^{-\frac{2\pi i}{48}} & 0 & 0\\
0 & e^{2\pi i\frac{23}{48}} & 0 \\
0 & 0 & e^{\frac{2\pi i}{24}}\end{pmatrix}\qquad S=\frac{1}{2}\begin{pmatrix}1 & 1 & \sqrt{2}\\
1 & 1 & -\sqrt{2}\\ \sqrt{2} & -\sqrt{2} & 0\end{pmatrix}\ .
\ee
The unique modular invariant function is the diagonal invariant
\bea
Z_{\rm AA}&=&\sum_{r,s}|\chi_{r,s}|^2
=|\chi_{1,1}|^2+|\chi_{1,2}|^2+|\chi_{2,1}|^2~.
\eea

\subsubsection*{Tricritical Ising Model}
The tricritical Ising model corresponds to the $(4,5)$ minimal model with central charge $c=\frac{7}{10}$. There are six primary fields with weights 
\be
0\ ,~\frac{1}{10}\ ,~\frac{3}{5}\ ,~\frac{3}{2}\ ,~\frac{3}{80}\ , ~\frac{7}{16}\ ,
\ee
and the basis of characters are
\be
\chi_{1,1}\ ,~\chi_{3,3}\ ,~\chi_{2,3}\ , ~\chi_{1,3}\ ,~\chi_{2,2}\ , ~\chi_{1,2}\ .
\ee
The representation of $SL_2(\ZZ)$ is given by
\be T=\diag(e^{2\pi i\frac{233}{240}},\,e^{2\pi i\frac{17}{240}},\,e^{2\pi i\frac{137}{240}},\,e^{2\pi i\frac{113}{240}},\,e^{2\pi i\frac{2}{240}},\,e^{2\pi i\frac{98}{240}})
\ee
\be S=\frac{1}{\sqrt{5}}\begin{pmatrix}
 s_2 & s_1 & s_1 & s_2 & \sqrt{2}s_1 & \sqrt{2}s_2\\
 s_1 & -s_2 & -s_2 & s_1 & \sqrt{2}s_2 & -\sqrt{2}s_1\\
 s_1 & -s_2 & -s_2 & s_1 & -\sqrt{2}s_2 & \sqrt{2}s_1\\
 s_2 & s_1 & s_1 & s_2 & -\sqrt{2}s_1 & -\sqrt{2}s_2\\
 \sqrt{2}s_1 & \sqrt{2}s_2 & -\sqrt{2}s_2 & -\sqrt{2}s_1 & 0&0\\
 \sqrt{2}s_2 & -\sqrt{2}s_1 & \sqrt{2}s_1 & -\sqrt{2}s_2 & 0&0
\end{pmatrix}\ ,
\ee 
where $s_1=\sin\frac{2\pi}{5}, s_2=\sin\frac{4\pi}{5}$. 
The unique modular invariant function is the diagonal invariant
\bea
Z_{\rm AA}&=&\sum_{r,s}|\chi_{r,s}|^2
= |\chi_{1,1}|^2+|\chi_{1,2}|^2+|\chi_{1,3}|^2+|\chi_{2,2}|^2+|\chi_{2,3}|^2+|\chi_{3,3}|^2~.
\eea

\subsubsection*{Three-State Potts Model}

The central charge of the $(5,6)$ model is $c=\frac{4}{5}$, there are $10$ characters 
\be
\chi_{1, 1}\ , ~ \chi_{1, 2}\ , ~ \chi_{1, 3}\ , ~ \chi_{1, 4}\ , ~ \chi_{2, 2}\ , ~ \chi_{2, 3}\ , ~ \chi_{2, 4}\ , ~ \chi_{3, 3}\ , ~ \chi_{3,4}\ , ~ \chi_{4, 4}~,\ee
corresponding to primaries of weight
\be
0\ ,~\frac{2}{5}\ ,~\frac{7}{5}\ ,~3\ ,~\frac{1}{40}\ ,~\frac{21}{40}\ ,~\frac{13}{8}\ ,~\frac{1}{15}\ ,~\frac{2}{3}\ ,~\frac{1}{8}~.\ee
There are two physical partition functions for $c={4\over 5}$. The diagonal invariant is
\be\label{inv1} Z_{\rm AA}=\sum_{r,s}|\chi_{r,s}|^2~,
\ee
and the block diagonal invariant is
\be\label{inv2} Z_{\rm DA}=|\chi_{1,1}+\chi_{1,4}|^2+|\chi_{1,2}+\chi_{1,3}|^2+2|\chi_{3,3}|^2+2|\chi_{3,4}|^2\ .
\ee
This block diagonal invariant is the partition function of the three-state Potts model.

\subsubsection*{Tricritical Potts Model}
Finally we turn to the (6,7) minimal model. The central charge is $c={6\over 7}$ and there are 15 irreducible characters $\chi_{r,s}$
\be
\chi_{1,1\cdots 5}\ , ~\chi_{2,2\cdots 5}\ , ~\chi_{3,3\cdots 5}\ ,~\chi_{4,4}\ ,~\chi_{4,5}\ ~\chi_{5,5}~,
\ee
with weights $h_{r,s}$
\be
0\ ,~\frac{3}{8}\ ,~\frac{4}{3}\ ,~\frac{23}{8}\ ,~5\ ,~\frac{1}{56}\ ,~\frac{10}{21}\ ,~\frac{85}{56}\ ,~\frac{22}{7}\ ,~\frac{1}{21}\ ,~\frac{33}{56}\ ,~\frac{12}{7}\ ,~\frac{5}{56}\ ,~\frac{5}{7}\ ,~\frac{1}{7}~.\ee
There are two physical partition functions for $c={6\over 7}$. The diagonal invariant is
\be\label{inv1t} Z_{\rm AA}=\sum_{r,s}|\chi_{r,s}|^2~,
\ee
and the block diagonal invariant is
\be\label{inv2t} Z_{\rm AD}=\sum_{s=1,2,3} |\chi_{s,1}+\chi_{s,5}|^2+2|\chi_{s,3}|^2\ .
\ee
This block diagonal invariant is the partition function of the tricritical Potts model.

\section{Modular Sums}\label{app:ms}

\subsection{Bases of modular invariants}\label{app:bmi}

In the $(p,p')$-minimal model, the primary fields are labeled by $(r,s)$, with $r=1,\ldots,p'-1$, 
$s=1,\ldots,p-1$ and the identification $(r,s)\sim (p'-r,p-s)$. The corresponding characters form 
a representation of $PSL(2,\ZZ)$ of dimension $(p-1)(p'-1)/2$, with generators $S$ and $T$ 
given by \eqref{app:aa} and \eqref{app:ab}. Modular invariant combinations of characters 
correspond to hermitian matrices which commute with $S$ and $T$. A basis for the space of such matrices 
has been found by Cappelli-Itzykson-Zuber \cite{Cappelli:1986hf,Cappelli:1987xt}. In this appendix, 
we will review their construction.

\bigskip

\noindent Let $p,p'$ be coprime positive integers.
For each $r,s\in \ZZ$, we define $\lambda_{r,s}\in\ZZ_{2pp'}$ by
\be \lambda_{r,s}:=rp-sp'\mod 2pp'\ .
\ee The following relations hold modulo $2pp'$
\be \lambda_{r,s}=-\lambda_{p'-r,p-s}=\omega_0\lambda_{r,-s}=-\omega_0\lambda_{-r,s}\mod 2pp'\ ,
\ee where $\omega_0$ is defined mod $2p p'$ by the properties
\be \omega_0p=p\mod 2pp'\ ,\qquad \omega_0p'=-p'\mod 2pp'\ ,\qquad \omega_0^2=1\mod 4pp'\ .
\ee More precisely, if $r_0,s_0$ are integers such that $r_0p-s_0p'=1$, then $\omega_0=r_0p+s_0p'$.
\smallskip

For $1\le r\le p'-1$ and $1\le s\le p-1$, the element $\lambda_{r,s}\in\ZZ_{2pp'}$ is related to the conformal weight $h_{r,s}$ of the $(r,s)$-primary field by 
\be 
h_{r,s}-\frac{c}{24}=\frac{\lambda_{r,s}^2}{4pp'}-\frac{1}{24}\mod 1\ ,
\ee
where $c\equiv c(p,p')$ is the central charge (\ref{cdef}). Furthermore, if we set
\be K_\lambda:=\frac{1}{\eta(\tau)}\sum_{n=-\infty}^{\infty}q^{\frac{(2pp'n+\lambda)^2}{4pp'}}\ ,
\qquad\text{ and }\qquad
\chi_{\lambda}:=K_{\lambda}-K_{\omega_0\lambda}\ ,
\ee 
then for $1\le r\le p'-1$ and $1\le s\le p-1$
\be K_{\lambda_{r,s}}= K_{r,s}\qquad\qquad\qquad \chi_{\lambda_{r,s}}=\chi_{r,s}\ ,
\ee
with $K_{r,s}$ and $\chi_{r,s}$ defined in \eqref{app:cfa}. The advantage of these definitions is that the modular transformations of $\chi_\lambda$ can be easily described as
\begin{align} &\chi_{\lambda}(\tau+1)=\sum_{\mu\in\ZZ_{2pp'}}T_{\lambda\mu}\chi_\mu(\tau)\ ,
&\chi_{\lambda}(-\tau^{-1})=\sum_{\mu\in\ZZ_{2pp'}}S_{\lambda\mu}\chi_\mu(\tau)\ ,
\end{align} where the $2pp'\times 2pp'$ matrices $T_{\lambda\mu}$ and $S_{\lambda\mu}$ are
\be\label{STdef} T_{\lambda\mu}=\delta^{(2pp')}(\lambda-\mu)\,e^{2\pi i\bigl(\frac{\lambda^2}{4pp'}-\frac{1}{24}\bigr)}\ ,\qquad \qquad
S_{\lambda\mu}=\frac{1}{\sqrt{2pp'}}e^{2\pi i\frac{\lambda\mu}{2pp'}}\ .
\ee
Using the identities
\be K_{\lambda}=K_{-\lambda}=K_{\lambda+2pp'}\ ,\qquad \qquad \chi_{\lambda}=\chi_{-\lambda}=\chi_{\lambda+2pp'}=-\chi_{\pm\omega_0\lambda}\ ,
\ee
it follows that each (not necessarily physical) modular invariant combination of characters of 
the $(p,p')$-minimal model corresponds to a $2pp'\times 2pp'$-dimensional hermitian matrix 
$X_{\lambda\mu}$ that commutes with $T_{\lambda\mu}$ and $S_{\lambda\mu}$ and satisfies
\be\label{Xconds} 
X_{\lambda,\mu}=X_{\lambda,-\mu}
=-X_{\lambda,\omega_0\mu}=-X_{\lambda,-\omega_0\mu}\ .
\ee
The invariant $Z(X)$ corresponding to such a $X_{\lambda\mu}$ is 
\be\label{Xmodinv} Z(X)=\frac{1}{16}\sum_{\lambda,\tilde\lambda\in \ZZ_{2pp'}} X_{\lambda,\tilde\lambda}\, \chi_\lambda(\tau)\bar\chi_{\tilde\lambda}(\tau)=\frac{1}{4}\sum_{\substack{1\le r,\tilde r \le p'-1\\1\le s,\tilde s \le p-1}}X_{\lambda_{r,s},\lambda_{\tilde r,\tilde s}}\,\chi_{r,s}(\tau) \bar\chi_{\tilde r,\tilde s}(\tau)\ .
\ee 

A basis for the space of $2pp'\times 2pp'$-dimensional real symmetric matrices commuting with 
$T_{\lambda\mu}$ and $S_{\lambda\mu}$ has been constructed by Cappelli, Itzykson and Zuber 
\cite{Cappelli:1987xt}. The basis elements
\be 
(\Omega^{pp'}_{dd'})_{\lambda,\mu}\ ,\qquad\qquad d,d'\in\ZZ,\quad d,d'>0,\quad d|p,\ d'|p'\ ,
\ee
have integral entries, and are in one to one correspondence with positive integers $d$ and $d'$ such that 
$d|p$ and $d'|p'$, see  \cite[eq.\ III.4]{Cappelli:1987xt}. The matrices 
$(\Omega^{pp'}_{dd'})_{\lambda,\mu}$ satisfy the relations
\be\label{omprops2} 
(\Omega^{pp'}_{dd'})_{\lambda,\mu}=(\Omega^{pp'}_{(p/d)d'})_{\lambda,\omega_0\mu}
=(\Omega^{pp'}_{d(p'/d')})_{\lambda,-\omega_0\mu}=(\Omega^{pp'}_{(p/d)(p'/d')})_{\lambda,-\mu}\ .
\ee 
The subspace of matrices satisfying the additional conditions \eqref{Xconds} is spanned by
\begin{align} X^{(d,d')}_{\lambda,\mu}:=&(\Omega^{pp'}_{dd'})_{\lambda,\mu}+(\Omega^{pp'}_{dd'})_{\lambda,-\mu}-(\Omega^{pp'}_{dd'})_{\lambda,\omega_0\mu}-(\Omega^{pp'}_{dd'})_{\lambda,-\omega_0\mu}\\=&(\Omega^{pp'}_{dd'})_{\lambda,\mu}+(\Omega^{pp'}_{(p/d)(p'/d')})_{\lambda,\mu}-(\Omega^{pp'}_{(p/d)p'})_{\lambda,\mu}-(\Omega^{pp'}_{d(p'/d')})_{\lambda,\mu}\ ,
\end{align} and the only linear relations are
\be X^{(d,d')}=-X^{(p/d,d')}=-X^{(d,p'/d')}=X^{(p/d,p'/d')}\ .
\ee

\subsection{Models with one or two modular invariants}\label{app:twoinvariants}

In this subsection we discuss the minimal models with only one or two modular invariants. In the 
first case, the sum over modular images of any seed, if not zero, must correspond to a physical modular 
invariant up to normalization. 

The $(p,p')$-minimal model has only one invariant $X^{(1,1)}$ if and only if both $p$ and $p'$ 
are either primes or squares of primes. Thus, there is an infinite set of minimal models --- with
the exception of the $(3,4)$- and $(4,5)$-models they are non-unitary --- for which the sum over 
the modular images of 
$|\chi_{1,1}|^2$ gives the AA physical invariant (up to the normalization)
\be 
\sum_{\gamma\in\Gamma} |\chi_{1,1}(\gamma\tau)|^2 \sim \sum_{r,s} |\chi_{r,s}(\tau)|^2\ .
\ee

The unitary minimal models $(m,m+1)$ with $m>4$ have always at least a second modular invariant. 
More generally, for any $(p,p')$-minimal model with $p>4$ even, there are exactly two modular invariants 
(namely, $X^{(1,1)}$ and $X^{(2,1)}$) if and only if $p'$ is prime or the square of a prime, and $p$ is 
twice a prime or $p=8$. In this case, with the exception of the models with 
$p=8$, one can define an extended chiral algebra with vacuum character 
$\chi_{1,1}+\chi_{p-1,1}$. The modular invariants $X^{(1,1)}$ and $X^{(2,1)}$ satisfy
\be 
\Tr\bigl((X^{(1,1)})^2\bigr)=\Tr\bigl((X^{(2,1)})^2\bigr)=(p-1)(p'-1)/2\ ,
\ee 
so that $I_1=X^{(1,1)}-X^{(2,1)}$ and $I_2=X^{(1,1)}+X^{(2,1)}$ form an orthogonal basis. 
Furthermore, if we define the hermitian matrix $Y$ by
\be 
\sum_{\lambda,\mu}Y_{\lambda\mu}\chi_\lambda\bar\chi_\mu=|\chi_{1,1}+\chi_{1,p-1}|^2\ ,
\ee 
we obtain
\be 
\Tr(I_1Y)=4\ ,\qquad \Tr(I_2Y)=0\ .
\ee 
Thus,
\be 
\sum_{\gamma\in\Gamma}\gamma\cdot Y\sim  I_1\ ,
\ee 
so that the sum over modular images gives (up to normalisation) the AD physical invariant. 

One way to generate unitary models of this kind is to consider a Sophie Germain prime  $p$, and
to set $m=2p$ with $m+1=2p+1$. (Sophie Germain primes are characterised by the property
that $2p+1$ is also prime.) It has been conjectured that there are infinitely many Sophie Germain
primes, see e.g.\ \cite{HW}, although this does not seem to have been proven yet.

\subsection{More than two invariants: the $(14,15)$ minimal model}\label{app:unphysical}

In general, when a model has many modular invariants, we do not expect the sum over the modular
images of the vacuum character to correspond to a physical invariant. 
A simple example is provided by the $(14,15)$ Virasoro minimal model. This model has four modular invariants
\be\label{invs1514} 
X^{(1,1)}\ ,\qquad X^{(2,1)}\ ,\qquad X^{(1,3)}\ ,\qquad X^{(2,3)}\ ,
\ee 
and the matrix of scalar products $(X^{(d,d')},X^{(\tilde d,\tilde{d}')}):=\Tr(X^{(d,d')}X^{(\tilde d,\tilde{d}')})$ is
\be 
\begin{pmatrix} 91 & 35 & 13 & 5\\ 35 & 91 & 5 & 13\\ 13 & 5 & 91 & 35\\ 5 & 13 & 35 & 91\end{pmatrix}\ ,
\ee 
where the invariants $X^{(d,d')}$ are ordered as in \eqref{invs1514}. An orthonormal basis is given by
\begin{align} 
I_1=&\frac{X^{(1,1)}-X^{(2,1)}}{\sqrt{112}}\ ,
& I_2=&\frac{X^{(1,1)}+X^{(2,1)}}{\sqrt{252}}\ ,\\ 
I_3=&\frac{-X^{(1,1)}+7X^{(1,3)}}{\sqrt{4368}}\ ,
& I_4=&\frac{-5X^{(1,1)}+13X^{(2,1)}+35X^{(1,3)}-91X^{(2,3)}}{\sqrt{628992}}\ .
\end{align}  
The physical modular invariants are $Z_{\rm AA}$ and $Z_{\rm DA}$, corresponding, 
up to normalization, to $X^{(1,1)}=\sqrt{28}I_1+\sqrt{63}I_2$ and $I_1$, respectively. 
Any linear combination of invariants involving $I_3$ and $I_4$ is non-physical. The invariant 
$Z_{\rm DA}$ is the partition function for a model with an extended chiral algebra, whose 
vacuum character is 
\be 
\chi_{1,1}+\chi_{1,13}\ .
\ee 
If we sum over the modular images of 
\be 
\sum_{\lambda,\mu}Y_{\lambda\mu}\chi_\lambda\bar\chi_\mu=|\chi_{1,1}+\chi_{1,13}|^2\ ,
\ee 
we obtain
\be\label{model1514} 
\sum_{\gamma\in\Gamma}\gamma\cdot Y
= |\Gamma| \Bigl(\frac{4}{\sqrt{112}}I_1-\frac{2}{\sqrt{4368}}I_3-\frac{36}{\sqrt{628992}}I_4\Bigr)\ .
\ee  
Thus, the modular invariant obtained this way is non-physical.  More generally, one can consider 
a seed of the form
\be 
\sum_{\lambda,\mu}Y_{\lambda\mu}\chi_\lambda\bar\chi_\mu=|\chi_{1,1}+\sum_{r,s} c_{r,s}\chi_{r,s}|^2\ ,
\ee 
where $c_{r,s}$ are non-negative integers. This seed corresponds to the vacuum character with 
respect to an extension of Virasoro algebra by $c_{r,s}$ copies of the $(r,s)$-fields. 
A necessary consistency condition for this algebra to be well defined is that $c_{r,s}=0$ whenever the 
conformal weight $h_{r,s}$ is not integral; for the $(14,15)$-model, the most general seed 
satisfying this constraint is
\be 
\sum_{\lambda,\mu}Y_{\lambda\mu}\chi_\lambda\bar\chi_\mu=|\chi_{1,1}+c_{1,13}\chi_{1,13}
+c_{4,13}\chi_{4,13}+c_{11,13}\chi_{11,13}|^2\ .
\ee 
The sum over modular images of this seed is proportional to the AA invariant 
$\sqrt{28}I_1+\sqrt{63}I_2$ if
\be\label{AAsol} 
\Tr(I_1Y)=\sqrt{28}K\ ,\qquad \Tr(I_2Y)=\sqrt{63}K\ ,\qquad \Tr(I_3Y)=0\ ,\qquad \Tr(I_4Y)=0\ ,
\ee 
and proportional to the AD invariant $I_1$ if
\be\label{ADsol} 
\Tr(I_1Y)=K\ ,\qquad \Tr(I_2Y)=0\ ,\qquad \Tr(I_3Y)=0\ ,\qquad \Tr(I_4Y)=0\ ,
\ee 
for an arbitrary real constant $K$. These equations can be solved for the unknowns 
$c_{1,13}$, $c_{4,13}$, $c_{11,13}$, and $K$, and one finds that there are no solutions 
where $c_{1,13}$, $c_{4,13}$, $c_{11,13}$ are all non-negative integers. Thus, no physical 
invariant can be obtained by seeds of this form. 

One could have also reached this conclusion by noting that if a consistent
extended algebra existed, then it would lead to another physical modular invariant, namely
the charge conjugation invariant with respect to this algebra. Since the classification of 
Cappelli-Itzykson-Zuber  \cite{Cappelli:1986hf,Cappelli:1987xt} excludes any further
invariant, it follows that there are only two consistent algebras, the Virasoro algebra
or its extension  by the $(r,s)=(1,13)$  field. In both cases one checks easily that the
modular completion does not lead to a physical invariant.

A natural way to relax the above ansatz is to consider  seeds containing primaries 
(with respect to the Virasoro or an extended algebra) other than the vacuum, but
to require that the corresponding conformal dimensions satisfy 
$h+\bar h\le \frac{c+\bar c}{24}$. By the argument above, we can restrict 
ourselves to the cases where the algebra is either the Virasoro algebra, or its 
extension by the $(r,s)=(1,13)$  field.  In the Virasoro case 
the most general seed is
\be 
\sum_{\lambda,\mu}Y_{\lambda\mu}\chi_\lambda\bar\chi_\mu=|\chi_{1,1}|^2+c_{2,2}|\chi_{2,2}|^2+c_{3,3}|\chi_{3,3}|^2
+c_{4,4}|\chi_{4,4}|^2+c_{5,5}|\chi_{5,5}|^2\ ,
\ee 
while for the extended algebra we have
\be 
\sum_{\lambda,\mu}Y_{\lambda\mu}\chi_\lambda\bar\chi_\mu=|\chi_{1,1}+\chi_{1,13}|^2+c_{3,3}
|\chi_{3,3}+\chi_{3,11}|^2+c_{5,5}|\chi_{5,5}+\chi_{5,9}|^2\ .
\ee
One can check that there are no solutions to \eqref{AAsol} and \eqref{ADsol} in the Virasoro case, 
while for the extended algebra the AD physical invariant is obtained from the seed
\be 
|\chi_{1,1}+\chi_{1,13}|^2+\frac{1}{6}|\chi_{3,3}+\chi_{3,11}|^2 \ .
\ee
However, the relative factor of $\tfrac{1}{6}$ means that this does not have an interpretation
as `the perturbative part' of a consistent partition function.


\begin{thebibliography}{99}


\bibitem{Witten:2007kt}
E.~Witten,
``Three-Dimensional Gravity Revisited,''
{\tt [arXiv:0706.3359 [hep-th]]}.

\bibitem{Maloney:2009ck}
A.~Maloney, W.~Song, A.~Strominger,
``Chiral Gravity, Log Gravity and Extremal CFT,''
Phys.\ Rev.\  {\bf D81}, 064007 (2010).
{\tt [arXiv:0903.4573 [hep-th]]}.
  
\bibitem{Gaberdiel:2010pz}
M.R.~Gaberdiel, R.~Gopakumar,
``An AdS$_3$ Dual for Minimal Model CFTs,''
Phys.\ Rev.\  {\bf D83}, 066007 (2011).
{\tt [arXiv:1011.2986 [hep-th]]}.

\bibitem{Dijkgraaf:2000fq}
R.~Dijkgraaf, J.M.~Maldacena, G.W.~Moore, E.P.~Verlinde,
``A Black hole Farey tail,''
{\tt [hep-th/0005003]}.
  
\bibitem{Maloney:2007ud}
A.~Maloney, E.~Witten,
``Quantum Gravity Partition Functions in Three Dimensions,''
JHEP {\bf 1002}, 029 (2010).
{\tt [arXiv:0712.0155 [hep-th]]}.
  
\bibitem{Brown:1986nw}
J.D.~Brown, M.~Henneaux,
``Central Charges in the Canonical Realization of Asymptotic Symmetries: An 
Example from Three-Dimensional Gravity,''
Commun.\ Math.\ Phys.\  {\bf 104}, 207 (1986).

\bibitem{Castro:2011ui}
A.~Castro, T.~Hartman, A.~Maloney,
``The Gravitational Exclusion Principle and Null States in Anti-de Sitter Space,''
{\tt [arXiv:1107.5098 [hep-th]]}.

\bibitem{Kraus:2006nb}
P.~Kraus, F.~Larsen,
``Partition functions and elliptic genera from supergravity,''
JHEP {\bf 0701}, 002 (2007).
{\tt [hep-th/0607138]}.

\bibitem{Manschot:2007ha}
J.~Manschot, G.W.~Moore,
``A Modern Farey Tail,''
Commun.\ Num.\ Theor.\ Phys.\  {\bf 4}, 103 (2010).
{\tt [arXiv:0712.0573 [hep-th]]}.

\bibitem{deBoer:2006vg}
J.~de Boer, M.C.N.~Cheng, R.~Dijkgraaf, J.~Manschot, E.~Verlinde,
``A Farey Tail for Attractor Black Holes,''
JHEP {\bf 0611}, 024 (2006).
{\tt [hep-th/0608059]}.


\bibitem{Strominger:2008dp}
A.~Strominger,
 ``A Simple Proof of the Chiral Gravity Conjecture,''
{\tt [arXiv:0808.0506 [hep-th]]}.
  
\bibitem{Cheng:2011ay}
M.C.N.~Cheng, J.F.R.~Duncan,
``On Rademacher Sums, the Largest Mathieu Group, and the Holographic Modularity of Moonshine,''
{\tt [arXiv:1110.3859 [math.RT]]}.  

\bibitem{Cappelli:1987xt}
A.~Cappelli, C.~Itzykson, J.B.~Zuber,
``The ADE Classification of Minimal and A1(1) Conformal Invariant Theories,''
Commun.\ Math.\ Phys.\  {\bf 113}, 1 (1987).
  
\bibitem{Blencowe:1988gj}
M.P.~Blencowe,
``A Consistent Interacting Massless Higher Spin Field Theory In D = (2+1),''
Class.\ Quant.\ Grav.\  {\bf 6}, 443 (1989).  
  
\bibitem{Campoleoni:2010zq}
A.~Campoleoni, S.~Fredenhagen, S.~Pfenninger, S.~Theisen,
``Asymptotic symmetries of three-dimensional gravity coupled to higher-spin fields,''
JHEP {\bf 1011}, 007 (2010).
{\tt [arXiv:1008.4744 [hep-th]]}.

\bibitem{Gaberdiel:2010jf}
M.R.~Gaberdiel, C.A.~Keller, R.~Volpato,
``Genus two partition functions of chiral conformal field theories,''
Commun.\ Num.\ Theor.\ Phys.\  {\bf 4}, 295 (2010).
{\tt [arXiv:1002.3371 [hep-th]]}.
  
\bibitem{Maldacena:1998bw}
J.M.~Maldacena, A.~Strominger,
``AdS(3) black holes and a stringy exclusion principle,''
JHEP {\bf 9812}, 005 (1998).
{\tt [hep-th/9804085]}.

\bibitem{Kraus:2006wn}
P.~Kraus,
``Lectures on black holes and the AdS(3) / CFT(2) correspondence,''
Lect.\ Notes Phys.\  {\bf 755}, 193 (2008).
{\tt [hep-th/0609074]}.
  
\bibitem{LR}
R.~Lawrence, L.~Rozansky, 
``Witten-Reshetikhin-Turaev Invariants of Seifert Manifolds,"
Comm.\ Math.\ Phys.\ {\bf 205}, 287 (1999).
  
\bibitem{Banados:1992wn}
M.~Banados, C.~Teitelboim, J.~Zanelli,
``The Black hole in three-dimensional space-time,''
Phys.\ Rev.\ Lett.\  {\bf 69}, 1849 (1992).
{\tt [hep-th/9204099]}.
  
  
 \bibitem{Castro:2010ce}
A.~Castro, A.~Lepage-Jutier, A.~Maloney,
``Higher Spin Theories in AdS$_3$ and a Gravitational Exclusion Principle,''
JHEP {\bf 1101}, 142 (2011).
{\tt  [arXiv:1012.0598 [hep-th]]}.

\bibitem{Witten:1987ty}
E.~Witten,
``Coadjoint Orbits of the Virasoro Group,''
Commun.\ Math.\ Phys.\  {\bf 114}, 1 (1988).
    
\bibitem{Aldaya:1989ra}
  V.~Aldaya, J.~Navarro-Salas,
``Quantization On The Virasoro Group,''
Commun.\ Math.\ Phys.\  {\bf 126}, 575 (1989).
  
\bibitem{DiFrancesco:1997nk}
P.~Di Francesco, P.~Mathieu, D.~Senechal,
``Conformal field theory,''
Springer (1997).
  
\bibitem{Achucarro:1987vz}
A.~Achucarro, P.K.~Townsend,
``A Chern-Simons Action for Three-Dimensional anti-De Sitter Supergravity Theories,''
Phys.\ Lett.\  {\bf B180}, 89 (1986).

\bibitem{Achucarro:1989gm}
A.~Achucarro, P.K.~Townsend,
``Extended Supergravities in d=(2+1) as Chern-Simons Theories,"
Phys.\ Lett.\  {\bf B229}, 383 (1989).
  
\bibitem{Aragone:1983sz}
C.~Aragone, S.~Deser, 
``Hypersymmetry in D = 3 of Coupled Gravity-Massless Spin-5/2 System,''
Class. Quant. Grav. {\bf 1}, L9 (1984).

\bibitem{Bergshoeff:1989ns}
E.~Bergshoeff, M.P.~Blencowe, K.S.~Stelle, 
``Area Preserving Diffeomorphisms and Higher Spin Algebra,'' 
Commun.\ Math.\ Phys.\  {\bf 128}, 213 (1990).

\bibitem{Banados:1998pi}
M.~Banados, K.~Bautier, O.~Coussaert, M.~Henneaux, M.~Ortiz,
``Anti-de Sitter / CFT correspondence in three-dimensional supergravity,''
Phys.\ Rev.\  {\bf D58}, 085020 (1998).
{\tt [arXiv:hep-th/9805165]}.

\bibitem{Henneaux:1999ib}
M.~Henneaux, L.~Maoz, A.~Schwimmer,
``Asymptotic dynamics and asymptotic symmetries of three-dimensional extended AdS supergravity,''
Annals Phys.\  {\bf 282}, 31 (2000).  {\tt [hep-th/9910013]}.
  
\bibitem{Henneaux:2010xg}
M.~Henneaux, S.-J.~Rey,
``Nonlinear $W_{\infty}$ as Asymptotic Symmetry of Three-Dimensional Higher Spin Anti-de Sitter Gravity,''
JHEP {\bf 1012}, 007 (2010).
{\tt [arXiv:1008.4579 [hep-th]]}.

\bibitem{Gaberdiel:2011wb}
M.R. Gaberdiel and T.~Hartman, 
``Symmetries of Holographic Minimal Models,''
JHEP {\bf 1105}, 031 (2011). {\tt [arXiv:1101.2910 [hep-th]]}.

\bibitem{Campoleoni:2011hg}
A.~Campoleoni, S.~Fredenhagen, S.~Pfenninger, 
``Asymptotic W-symmetries in three-dimensional higher-spin gauge theories,''
{\tt [arXiv:1107.0290 [hep-th]]}.

\bibitem{Bantay:2001ni}
P.~Bantay,
``The kernel of the modular representation and the Galois action in RCFT,''
Commun.\ Math.\ Phys.\  {\bf 233}, 423 (2003).
{\tt [math/0102149]}.

\bibitem{Cappelli:1986hf}
A.~Cappelli, C.~Itzykson, J.-B.~Zuber,
``Modular Invariant Partition Functions in Two-Dimensions,''
Nucl.\ Phys.\  {\bf B280}, 445 (1987).
  
\bibitem{Gannon:1994km}
T.~Gannon, M.A.~Walton,
``On the classification of diagonal coset modular invariants,''
Commun.\ Math.\ Phys.\  {\bf 173}, 175 (1995).
{\tt [hep-th/9407055]}.
 
\bibitem{Witten:1988hc}
E.~Witten,
``(2+1)-Dimensional Gravity as an Exactly Soluble System,''
Nucl.\ Phys.\  {\bf B311}, 46 (1988).  
    
\bibitem{Vasiliev:1989qh}
M.A.~Vasiliev,
``Quantization on sphere and high spin superalgebras,''
JETP Lett.\  {\bf 50}, 374 (1989).  
  
\bibitem{Fradkin:1986ka}
E.S.~Fradkin, M.A.~Vasiliev,
``Candidate to the Role of Higher Spin Symmetry,''
Annals Phys.\  {\bf 177}, 63 (1987).
  
\bibitem{Vasiliev:1986qx}
M.A.~Vasiliev,
``Extended Higher Spin Superalgebras And Their Realizations In Terms Of Quantum Operators,''
Fortsch.\ Phys.\  {\bf 36}, 33 (1988).
  
\bibitem{Vasiliev:2001ur}
M.A.~Vasiliev,
``Progress in higher spin gauge theories,''   {\tt [hep-th/0104246]}.  
  
\bibitem{Gutperle:2011kf}
M.~Gutperle, P.~Kraus,
``Higher Spin Black Holes,''
JHEP {\bf 1105}, 022 (2011).
{\tt [arXiv:1103.4304 [hep-th]]}.
  
\bibitem{Ammon:2011nk}
M.~Ammon, M.~Gutperle, P.~Kraus, E.~Perlmutter,
``Spacetime Geometry in Higher Spin Gravity,''
{\tt [arXiv:1106.4788 [hep-th]]}.  

\bibitem{Castro:2011fm}
A.~Castro, E.~Hijano, A.~Lepage-Jutier, A.~Maloney,
``Black Holes and Singularity Resolution in Higher Spin Gravity,''  {\tt [arXiv:1110.4117 [hep-th]]}.
  
\bibitem{Drinfeld:1984qv}
V.G.~Drinfeld, V.V.~Sokolov,
``Lie algebras and equations of Korteweg-de Vries type,''
J.\ Sov.\ Math.\  {\bf 30}, 1975 (1984).  
  
\bibitem{Fateev:1987zh}
V.A.~Fateev, S.L.~Lukyanov,
``The Models of Two-Dimensional Conformal Quantum Field Theory with Z(n) Symmetry,''
Int.\ J.\ Mod.\ Phys.\  {\bf A3}, 507 (1988).
  
\bibitem{Bouwknegt:1992wg}
P.~Bouwknegt, K.~Schoutens,
``W symmetry in conformal field theory,''
Phys.\ Rept.\  {\bf 223}, 183 (1993). {\tt [hep-th/9210010]}.

\bibitem{Gaberdiel:2011zw}
M.R.~Gaberdiel, R.~Gopakumar, T.~Hartman, S.~Raju,
``Partition Functions of Holographic Minimal Models,''
JHEP {\bf 1108}, 077 (2011).
{\tt [arXiv:1106.1897 [hep-th]]}.  
  
\bibitem{Verlinde:1989ua}
H.L.~Verlinde,
``Conformal Field Theory, 2-d Quantum Gravity And Quantization Of Teichmuller Space,''
Nucl.\ Phys.\  {\bf B337}, 652 (1990).

\bibitem{Coussaert:1995zp}
O.~Coussaert, M.~Henneaux, P.~van Driel,
``The Asymptotic dynamics of three-dimensional Einstein gravity with a negative cosmological constant,''
Class.\ Quant.\ Grav.\  {\bf 12}, 2961 (1995).
{\tt [gr-qc/9506019]}.  
  
\bibitem{Bouwknegt:1988sv}
P.~Bouwknegt,
``Extended Conformal Algebras,''
Phys.\ Lett.\  {\bf B207}, 295 (1988).
  
\bibitem{Gannon:2001py}
T.~Gannon,
``Algorithms for affine Kac-Moody algebras,''
{\tt [hep-th/0106123]}.  

\bibitem{Yin:2007gv}
X.~Yin,
``Partition Functions of Three-Dimensional Pure Gravity,''
{\tt [arXiv:0710.2129 [hep-th]]}.
  
\bibitem{Yin:2007at}
X.~Yin,
``On Non-handlebody Instantons in 3D Gravity,''
JHEP {\bf 0809}, 120 (2008).
{\tt [arXiv:0711.2803 [hep-th]]}.

\bibitem{Strominger:1997eq}
A.~Strominger,
``Black hole entropy from near-horizon microstates,''
JHEP {\bf 9802}, 009 (1998).
{\tt [arXiv:hep-th/9712251]}.

\bibitem{Kato:1987td}
A.~Kato,
``Classification Of Modular Invariant Partition Functions In Two-dimensions,''
Mod.\ Phys.\ Lett.\  {\bf A2}, 585 (1987).

\bibitem{Gannon:1999}
T.~Gannon,
``The Cappelli-Itzykson-Zuber A-D-E classification,''
Rev.\ Math.\ Phys.\ {\bf 12}, 82 (2000).  {\tt [arXiv:math/9902064]}.

\bibitem{HW}
G.H.~Hardy, E.M.~Wright, 
``An introduction to the theory of numbers," 
Oxford University Press (1979).




\end{thebibliography}
\end{document}